\documentclass[12pt, a4paper]{article}

\usepackage{amsmath,amsfonts,bigints} 
\usepackage{fullpage}
\usepackage{color}
\usepackage{tikz}
\usepackage{tkz-euclide}
\usepackage{url}
\usepackage{authblk}
\usepackage{graphicx} 
\usepackage{hyperref}
\usepackage[left=1.5cm, right=1.5cm, top=1.785cm, bottom=2.0cm]{geometry}

\usepackage{bm}
\usepackage{tikz}
\usepackage{subcaption}
\usepackage{amssymb}
\usepackage{mathtools}
\usepackage{siunitx}
\usepackage{multirow}
\usepackage{array}
\usepackage{setspace}

\usepackage[english]{babel}
\usepackage[square,numbers]{natbib}
\definecolor{cream}{RGB}{222,217,201}

\title{Uniaxial poroelastic tendon model with crimped fibre recruitment reveals hysteresis}

\author[1]{Zoe C.~Godard\footnote{Email: {\tt zoe.godard@maths.ox.ac.uk}}}
\author[1]{Sarah L.~Waters}
\author[1]{Derek E.~Moulton}
\affil[1]{Mathematical Institute, University of Oxford, Oxford, UK}
\date{}

\begin{document}

\maketitle

\onehalfspacing

\begin{abstract}
\onehalfspacing
Fibre (or fibril) recruitment plays an important role in tendon and other biological soft tissue mechanics. Due to their large water content, a popular modelling approach for tendons is poroelasticity. Within this framework some tendon studies have included fibres, though none have included crimped fibre recruitment. We present a one dimensional poroelastic model in which the solid skeleton is composed of a soft neo-Hookean background matrix (NCM) and crimped fibrils which do not bear load (FIB model). As the tissue is stretched, fibrils are straightened  and contribute to load bearing. The fibre-reinforced tissue is compared to a tissue with a purely neo-Hookean (NH) skeleton in response to a uniaxial constant applied load (loading) and release of the load (unloading). The system dynamics are governed by a diffusion equation where the diffusion coefficient depends on stiffness. Within tendon parameter ranges, the FIB model is softer than the NH model, and so approaches steady state more slowly during loading. The presence of stiff, crimped fibrils increases bulk extension without excessively straining the fibrils, providing a natural protection mechanism for the tendon's structural components to load, in agreement with experiments. During unloading, the FIB model is much slower to relax as the tissue softens due to fibril re-crimping. This asymmetry in loading and unloading manifests as a hysteresis loop in the stress-strain curve averaged over the tendon. The hysteresis is reduced with increasing applied load. To the authors' knowledge, the finding of hysteresis within this framework is novel. The inclusion of fibrils allows for clearer biological interpretation and potential comparison to data. While the stress law employed in this study is bespoke for the application at hand by accounting for crimp and fibril recruitment, other fibril constitutive laws can readily be considered and incorporated into this framework.
\end{abstract}

\newpage

\section{Introduction}

Tendons are connective soft tissues which play an important role in bearing load between muscle and bone. The link between tendon structure and mechanical function is complex and, despite numerous experimental and theoretical studies, is still not fully understood. Mathematical modelling offers a cheap and fast yet rich alternative to experimental studies to explore various hypotheses relating to tendon's behaviour. 

\begin{figure}
    \centering
    \includegraphics[width=7cm]{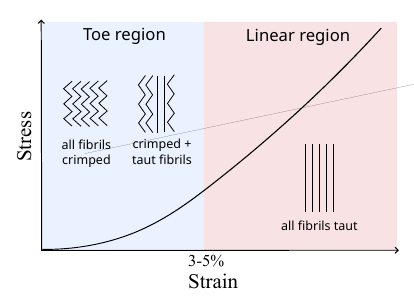}
    \caption{Diagram of tendon stress-strain relationship, as reported in \citet{thorpe:2015}. A popular theory is that the ``uncrimping'' generates the toe region observed in tendon's stress-strain relation \citep{kastelic:1980,screen:2004,screen:2004a,shearer:2015a}. In this model, the toe region is associated with crimped and uncrimping (crimped+taut) fibrils, while the linear region is associated with fully taut fibrils.}
    \label{fig:tendon-stress-strain}
\end{figure}

Tendons are primarily composed of collagen and water, as well as some cells and other non-collagenous components such as proteoglycans. Interestingly, their reported stress-strain relationship is non-linear for small deformations, referred to as the ``toe region'', and linear for large deformations (the ``linear region'') \citep{thorpe:2015}, as shown in Figure \ref{fig:tendon-stress-strain}. The main load-bearing component of tendon is the collagen hierarchical structure, which is axially aligned. It is well documented that collagen fibrils possess a wavy-like structure (crimp) and are ``uncrimped'' during loading \citep{diamant:1972,kastelic:1978}, only bearing load once uncrimped (or ``taut''). A popular theory is that the ``uncrimping'' generates the toe region observed in tendon's stress-strain relation \citep{kastelic:1980,screen:2004,screen:2004a,shearer:2015a}, and this feature explicitly appears in several mechanical models. For example, in \citep{shearer:2015a}, Shearer derives a strain-energy function for a fascicle (tendon subunit) made up of crimped collagen fibrils which only bear load once taut and are embedded in a soft background matrix. The parameters of the model are measurable, and Shearer finds improved experimental fit compared to the Holzapfel-Gasser-Ogden model which is commonly used for modelling biological tissues and requires phenomenological parameters. \citet{ciarletta:2008} also incorporates fibril recruitment into their viscoelastic model of tendon to partially explain hysteresis observed in tendon loading experiments \citep{screen:2004,ciarletta:2008,maganaris:2000}.

The above models only consider the solid components of tendon; due to its 55-70\% water content, poroelastic modelling is a popular approach to capture both the fluid and solid phases of tendon. For example, \citet{wren:2000} employ a finite-element poroelastic model to explore the impact of permeability in fibrocartilaginous regions of tendon, and \citet{safa:2020} investigate the transverse poroelastic properties of tendon using a biphasic finite element model. These studies are among numerous other linear and non-linear poroelastic models of tendon which examine properties such as Poisson's ratio \citep{ahmadzadeh:2015}, permeability \citep{chen:1998}, heterogeneity \citep{godard:2025} and interstitial fluid flow \citep{butler:1997}, or explore different modelling approaches such as the inclusion of viscoelasticity \citep{khayyeri:2015} or cells \citep{lavagnino:2008}. 

Some poroelastic models of tendon also include fibres. For example, \citet{atkinson:1997} present a cylindrical poroelastic model with a ring of helically arranged fibres on the outside, and \citet{khayyeri:2015} include viscoelastic fibres in the solid matrix of their poroelastic model, though neither study allows for fibril recruitment. Other studies investigate the fibre-reinforced poroelastic modelling framework in application to cartilage \citep{chen:2020,gupta:2010a}. More recently, \citet{hennessy:2025} employs a fibre-reinforced hydrogel model to capture cartilage response to compression, with an empirical fibre recruitment rule. Their study highlights the importance of accounting for recruitment when modelling soft tissues such as cartilage, ligament and tendon.

Within the poroelastic framework there exists an inherent timescale so that the response to load imposed at one end is not instantaneous in the rest of the tissue. The rate at which the response propagates through the tissue is governed by a diffusion coefficient which varies with porosity and stiffness. Studies have characterised the mechanics and flow of the non-linear poroelastic material in response to constant and cyclic loading \citep{macminn:2016,fiori:2023}. In our previous study, \citet{godard:2025}, we were motivated by how damage affects tendon's response to cyclic loading. The one-dimensional non-linear poroelastic model was built following \citet{macminn:2016}, and ``damage'' was incorporated as heterogeneous stiffness and permeability. The evolution of diffusion and porosity are tightly linked in non-linear poroelasticity and the results from our analysis highlighted the tissue’s sensitivity to stiffness via the diffusion coefficient. While our modelling framework was partly motivated by tendon, a compressible neo-Hookean law was employed to model the behaviour of the solid skeleton, and fibres were not included in the mechanical description. This raises the question: how does fibre recruitment in a non-linear poroelastic framework affect the tissue's response to loading and unloading? 

To address this, and motivated by the uncrimping behaviour observed experimentally, in this study we extend our previous one-dimensional poroelastic framework to incorporate fibre recruitment, assuming that the solid structure is made up of initially crimped collagen fibrils embedded in a Non-Collagenous Matrix (NCM) and subject to a uniaxial constant load. In this work, `fibril' refers to the tendon collagen subunit explicitly included in the model, but the term `fibre' is also employed in reference to `fibre recruitment' and `fibre-reinforcement', in accordance with poroelasticity literature.

In Section \ref{sec: model}, the fibril component of the solid matrix stress is derived following \citet{shearer:2015a} and adapted for uniaxial poroelasticity. This enables us to directly compare the response under load of the fibre-reinforced model (denoted the FIB model) and our previous model, a purely neo-Hookean solid skeleton (NH model). Although our poroelastic model is effectively one-dimensional, it accounts for the presence of many fibrils with a range of ``resting” (initial) crimp angles, so that the fibrils are not all simultaneously taut at a given axial position. In addition, due to the diffusive nature of the system, fibril recruitment is not uniform throughout the material’s length and depends on the axial coordinate. The parameter choices for the FIB and NH models are carefully discussed in relation to known tendon measurements. 

In Section \ref{sec: results}, we investigate two separate boundary conditions at the muscle-tendon interface (see Figure \ref{fig:diagram-tendon-fibrils}): loading, where the tissue begins at rest and is instantaneously loaded with a constant stress, and unloading, where the tissue begins in a loaded state and the load is instantaneously released. These conditions are motivated by typical experimental mechanical tests performed on tendons but also chosen for their mathematical simplicity, allowing for clear comparison between the models whilst uncovering information about the material's mechanics. Key differences between the FIB and NH responses to loading and unloading, and between the loading and unloading responses, are presented, as well as a more detailed analysis of the FIB model, including an analysis of hysteresis, a novel finding within this framework. Finally, in Section \ref{sec: discussion} we discuss the importance of our results within the poroelastic and biological context, and conclude in Section \ref{sec: conclusion}.

\section{Model}\label{sec: model}

\begin{figure}[h]
    \centering
    \includegraphics[width= 10cm]{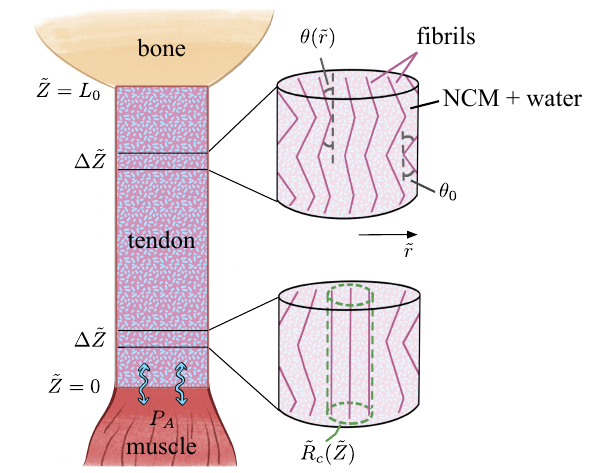}
    \caption{Poroelastic tendon bounded by muscle at $\tilde Z=0$ and bone at $\tilde Z=L_0$, with ambient fluid pressure $P_A$ at the muscle boundary. For a given element $\Delta \tilde Z$ we envisage a cylinder containing crimped fibrils embedded in NCM. Crimp angle $\theta(\tilde r)$ increases radially and, as the tendon is locally stretched, fibrils contained in the radius $\tilde R_c(\tilde Z)$ are uncrimped.}
    \label{fig:diagram-tendon-fibrils}
\end{figure}

Consider the deformation of a fully saturated poroelastic material in the Lagrangian domain
\begin{equation}0\leq \tilde{Z}\leq L_0,\end{equation}
where $L_0$ is the resting length of the tendon and $\tilde Z$ is the material (Lagrangian) axial coordinate, as shown in Figure \ref{fig:diagram-tendon-fibrils}. Tilde denotes dimensional variables. As we do not consider formal homogenisation, the solid skeleton in the poroelastic setting is comprised of all solid components in the tendon, primarily collagen, while the fluid-filled space between comprises the pores. Assume the boundaries of the system coincide with the boundaries of the solid; the tendon is bounded by muscle at $\tilde Z=0$ and bone at $\tilde Z=L_0$. Supposing that the material begins at time $\tilde t=0$ in a stress free state, the reference or initial configuration refers to the material's configuration at time $\tilde t=0$, while the current configuration refers to the material's configuration at current time $\tilde t$.  The model is presented in a Lagrangian framework, consistent with \citet{godard:2025}, meaning the nominal porosity $\Phi$ (current fluid volume fraction with respect to total initial volume) and nominal effective stress $\tilde s'$ (current force with respect to initial area) are considered, rather than the true (Eulerian) porosity $\phi$ and effective stress $\tilde\sigma'$ which are measured with respect to total current volume and current area respectively. Note we refer to the effective stress here rather than total stress, which accounts for the fact that fluid, which has fluid pressure $P$, exerts pressure on the solid skeleton even when at rest which does not contribute to solid deformation. Both the solid grains forming the solid skeleton and the fluid are assumed incompressible, so that when the poroelastic material is pulled axially, the pore size must increase (fluid enters), i.e.\ the solid fraction decreases as the total volume increases and vice versa. The fully saturated poroelastic material is therefore compressible through pore rearrangement. Fluid flows with pressure $P$, which is isotropic, through the material which is assumed to have a porosity-dependent permeability $k(\Phi)$. In addition, the permeability of the material $k(\Phi)$ is assumed to be porosity-dependent.

Tendons are highly anisotropic, with a hierarchical structure which is axially aligned, and are subject to loads that are primarily axial \cite{thorpe:2015}. For simplicity, in this model we adopt a uniaxial assumption, restricting attention to axial stress and strain only. This represents a reduced description of the tendon mechanics that does not resolve transverse deformation or stresses. The model should therefore be interpreted as a simplified framework intended to elucidate qualitative poroelastic response and the impact of fibre recruitment.
%it is assumed that deformation is uniaxial; in order for this to be possible while allowing the solid to stretch axially (as its pores expand), it is implicitly assumed that the transverse boundaries are fixed to a slippery wall, such that there is no transverse deformation but the solid can still slide along the transverse walls. This generates a transverse stress at the boundaries, which is not considered in this study since only the axial components of the stress and strain fields are investigated. The present formulation should therefore be interpreted as a reduced model intended to elucidate the qualitative axial poroelastic response rather than to resolve the complete three-dimensional stress and deformation state. 
Under this assumption, the deformation gradient tensor is given by
\begin{equation}\label{eq. F 1D}
    \mathbf{F} = \frac{\partial \mathbf{\tilde x}(\mathbf X,t)}{\partial \mathbf{\tilde X}} = \begin{pmatrix}
        1&0&0\\
        0&1&0 \\
        0&0&J(\tilde Z,\tilde t)
    \end{pmatrix}
\end{equation}
where $\mathbf{X}=(\tilde X,\tilde Y,\tilde Z)$ and $\mathbf{x}=(\tilde x,\tilde y,\tilde z)$ denote Lagrangian and Eulerian Cartesian coordinates respectively, and $J$ is the Jacobian determinant defined by
\begin{equation}\label{eq.Jacobian}
    J=\det(\mathbf{F}),
\end{equation}
representing volume change. For uniaxial deformation, $J$ is equivalent to stretch. Note that scalar quantities presented here correspond to the $zz-$ ($z-$) component of their respective tensorial (vector) fields, for example $s(Z,t)$ is the $zz-$component of $\mathbf{s(Z,t)}$ and $Q$ is the axial component of flux $\mathbf{Q}$.

At each $\tilde Z$, and for a given element $\Delta\tilde Z$, we envisage a cylinder containing a fraction $c_f$ of stiff, crimped fibrils embedded in a soft background matrix, the Non-Collagenous Matrix (NCM), as shown in Figure \ref{fig:diagram-tendon-fibrils}. The fibril fraction $c_f$ is uniform throughout the tendon. The fibrils are modelled as extensible, one-dimensional load-bearing elements whose axial deformation is determined by the deformation of the solid in which they are embedded. They contribute to the axial stress but are not assigned an independent three-dimensional constitutive response; consequently, their transverse deformation is not resolved, and no fibril Poisson's ratio enters the model. The local stress $\tilde s'(\tilde Z)$ is then the average contribution from the fibrils and the NCM to stress in the cylinder. Letting $0\leq \tilde r\leq R_0$ denote radial coordinates with $R_0$ the radius of the tendon, the fibrils are crimped according to the crimp angle distribution $\theta(\tilde r)$ and fibrils are assumed to only bear load once straightened. Fibril crimp is assumed to increase with $\tilde r$, generalising \citet{kastelic:1978}'s experimental observation of increasing crimp with radius in single tendon fascicles (a collagen sub-unit of tendon). As the tissue is locally stretched, fibrils contained in $0\leq \tilde r \leq \tilde R_c(\tilde Z)$ are taut and bear load, while fibrils outside $\tilde R_c(\tilde Z)$ remain crimped and bear no load; this corresponds to the toe region. Once all fibrils are taut, then $\tilde R_c(\tilde Z)=R_0$ so that all fibrils locally bear load; this is the linear region. Denoting $\tilde s_f$ and $\tilde s_m$ as the local average stress contribution from the fibrils and the NCM respectively, the nominal effective stress of the solid skeleton at $\tilde Z$ and at time $\tilde t$ is given by
\begin{equation}\label{eq. s' = sf + sm}
    \tilde s'(\tilde Z,\tilde t) = c_f\tilde s_f(\tilde Z,\tilde t)+(1-c_f)\tilde s_m(\tilde Z, \tilde t),
\end{equation}
where the fibril and NCM contributions to the effective stress are weighted by $c_f$ and $(1-c_f)$ respectively. The pointwise contribution to stress in $\tilde r$ from the fibrils and from the NCM is collapsed to one point for each $\tilde Z$. The expressions for $\tilde s_f$ and $\tilde s_m$ remain to be determined.

\subsection{Solid skeleton stress}\label{subsec. stress law}

We now derive expressions for $\tilde s_f$ and $\tilde s_m$. For $\tilde s_f$, consider the local stretch which is given by $J(\tilde Z,t)$. The recruitment stretch $J_c$ required to straighten a fibril at $\tilde r$ is related to crimp angle $\theta(\tilde r)$ by:
\begin{equation}\label{eq. recruitment stretch}
    J_c(\theta(\tilde r)) = \sec\theta(\tilde r).
\end{equation}
Letting $\theta_0=\theta(R_0)$ be the maximum crimp angle, the maximum recruitment stretch $J_c(\theta_0)$ required to straighten all fibrils
\begin{equation}\label{eq. max recruitment stretch}
    J_c(\theta(R_0)) = J_c(\theta_0) = \sec\theta_0.
\end{equation}
The associated strain $\Phi_c(\theta_0)=J_c(\theta_0)-1$ defines the limit between the toe and linear region of the stress-strain relationship. For a given stretch $1<J(\tilde Z,\tilde t)<J_c(\theta_0)$, the local radius of taut fibrils $\tilde R_c(\tilde Z,\tilde t)$ can be found by solving $J=J_c(\tilde R_c)$ which rearranges to
\begin{equation}\label{eq. Rc equation}
    \theta(\tilde R_c) = \arccos(1/J).
\end{equation}
For $J(\tilde Z,\tilde t)\geq J_c(\theta_0)$ then $\tilde R_c(\tilde Z,\tilde t)=\tilde R_c^{max}=R_0$.

Assuming each individual fibril is Hookean, \citet{shearer:2015a} derives the average Cauchy stress contribution from the fibrils $\tilde\sigma_f$ (Eulerian stress, with respect to current configuration) for a purely elastic cylinder made of crimped fibrils and soft background matrix, by relating the tendon (macroscopic) stretch $J$ to the individual fibril strain $e_f(\tilde r)=J\cos\theta(r)-1$, and integrating the individual fibril stress $\tilde s_f(\tilde r)=E_f e_f(\tilde r)$ over the area containing taut fibrils:
\begin{equation}
    \tilde\sigma_f(\tilde Z,t) = \frac{2E_f}{R_0^2}\begin{cases}
        \int_0^{\tilde R_c(\tilde Z,\tilde t)} [J(\tilde Z,\tilde t)\cos\theta(\tilde r)-1] \;\tilde rd\tilde r \quad\text{for}\quad 1\leq J(\tilde Z,\tilde t)<J_c(\theta_0) \\
        \int_0^{R_0} [J(\tilde Z,\tilde t)\cos\theta(\tilde r)-1] \;\tilde rd\tilde r \quad\text{for}\quad J(\tilde Z,\tilde t)\geq J_c(\theta_0)
        \end{cases}
\end{equation}
where $E_f$ is the fibrils' Young's modulus which has dimensions of force per area. Similarly, we can derive the average fibril strain $e_f(\tilde Z,t)$, by integrating the individual fibril strain $e_f(\tilde r)$ over the area containing taut fibrils, giving:
\begin{equation}
    e_f(\tilde Z,t) = \tilde\sigma_f(\tilde Z,t)/E_f.
\end{equation}
The Piola-Kirchoff (PK) stress (i.e.\ Lagrangian stress) is related to the Cauchy stress by $\mathbf{\tilde s} = J \mathbf{\tilde\sigma}\mathbf{F}^{-T}$ for a compressible solid. For uniaxial deformation, using \eqref{eq. F 1D}, the average PK stress contribution from the fibrils is then
\begin{equation}
    \tilde s_f(\tilde Z,\tilde t) = J\tilde\sigma_f J^{-1} = \tilde\sigma_f.
\end{equation}
A compressible neo-Hookean law \citep{ogden:2013,ahmadi:2025} is chosen to model the NCM:
\begin{equation}
    \tilde s_m(\tilde Z,\tilde t) = 2\biggl[C_1^m(J(\tilde Z,\tilde t)-\frac{1}{J(\tilde Z,\tilde t)})+D_1^m(J(\tilde Z,\tilde t)-1)\biggl]
\end{equation}
where the NCM coefficients are given by $C_1^m = E_m/4(1+\nu)$ and $D_1^m = E_m/6(1-2\nu)$, where $E_m$ is the NCM's Young's modulus and $\nu$ is the NCM's Poisson ratio.

Using the crimp angle distribution $\theta(\tilde r)=\sin^{-1}(\tilde r/R_0 \sin\theta_0)$ (following \citet{shearer:2015a}), the radius of taut fibrils for $1\leq J(\tilde Z, \tilde t) <J_c(\theta_0)$ is 
\begin{equation}\label{eq. Rc}
    \tilde R_c(\tilde Z,\tilde t) = R_0\sqrt{1-J^{-2}(\tilde Z,\tilde t)}/\sin\theta_0
\end{equation}
and the stress law becomes:
\begin{equation}\label{eq. effective stress uncrimp sin law}
    \tilde s'_{FIB} = 2(1-c_f)\biggl[C_1^m(J-\frac{1}{J})+D_1^m(J-1)\biggl] + c_fE_f\begin{cases}
        \frac{1}{3\sin^2\theta_0}(2J-3+J^{-2}) \quad\text{for}\quad 1\leq J<J_c(\theta_0) \\
        \frac{2(1-\cos^3\theta_0)}{3\sin^2\theta_0}J-1 \quad\text{for}\quad J>J_c(\theta_0)
    \end{cases}
\end{equation}
where subscript FIB denotes the FIB stress law. Note that once all the fibrils in $\tilde r$ at a given $\tilde Z$ are taut, the fibril contribution to the load is linear in $J(\tilde Z,\tilde t)$, where the scaling $2(1-\cos^3\theta_0)/3\sin^2\theta_0$ captures the fact that, locally, fibrils are not recruited at the same time. This is referred to as the FIB model.
% Note this set up means at say $\tilde Z=0.2$ all fibrils could be taut but at $\tilde Z=0.3$ they're not.

This will be compared to the neo-Hookean stress law (or ``NH'' for short), given by:
\begin{equation}\label{eq. effective stress neo}
    \tilde s'_{NH} = 2C_1^T(J-\frac{1}{J})+2D_1^T(J-1)
\end{equation}
where this time $C_1^T =E_T/4(1+\nu)$ and $D_1^T = E_T/6(1-2\nu)$ where $E_T$ is the tendon's Young's modulus as a whole. As reporting on Poisson's ratio in tendons is mixed, it is taken to be the same for the NCM and the whole tendon. This is referred to as the NH model. 

Throughout the paper, subscript FIB and NH is adopted to differentiate between FIB model and NH model quantities.

\subsection{Non-dimensionalisation}

Non-dimensionalisation is as follows:
\begin{equation}\label{eq. non-dim}
    Z = \tilde{Z}/L_0 \quad r = \tilde r/R_0 \quad R_c = \tilde R_c/R_0 \quad  s'= \tilde{s}'/E_f \quad t = \tilde t/ T_{pe} \quad k=\tilde k/k_0 \quad P=(\tilde P+P_A)/E_f,
\end{equation}
where $k_0$ is the reference permeability of the tissue, $P_A$ is ambient pressure (in this model it is the pressure at the muscle boundary) and $T_{pe}=\mu L_0^2/k_0E_f$ is the poroelastic timescale, where $\mu$ is the fluid's dynamic viscosity. Stress is chosen to be non-dimensionalised by the fibril Young's modulus $E_f$ since the collagen units are the main load-bearing components of tendon \citep{thorpe:2015}. The parameters $k_0$ and $P_A$ are presented here for completeness but are not discussed further as they are not needed for the non-dimensional model. The poroelastic timescale $T_{pe}$ characterises the time it takes for the response to load to propagate through the tissue. Again, this is relevant to the dimensional model but does not appear in the non-dimensional model and so is not discussed further. Note that $R_0$ does not appear in the non-dimensional model either (c.f.\ eq.\ \eqref{eq. Rc} with eq.\ \eqref{eq. non-dim}).

\subsection{Uniaxial poroelasticity}\label{subsec. uniaxial poroelasticity}

Here the key ingredients of the non-dimensional Lagrangian non-linear one-dimensional poroelastic framework are presented. The interested reader is referred to \citet{godard:2025} for more details, where the model was built following \citet{macminn:2016} and \citet{coussy:2004}. 

Let $U(Z,t) = z(Z,t)-Z$ denote the solid displacement. Combining with \eqref{eq. F 1D}:
\begin{equation}\label{eq. J = 1+U_Z}
    J(Z,t) = 1+ U_Z
\end{equation}
where $U_Z=\partial U/\partial Z$ is the solid strain. The solid grains forming the solid skeleton are incompressible; the incompressibility condition for the solid phase is given by
\begin{equation}\label{eq. Jacphi}
    J(Z,t) = 1 + \Phi(Z,t) - \Phi_0 \equiv 1+\hat{\Phi}(Z,t),
\end{equation}
where $\Phi(Z,t)$ is the nominal porosity, $\Phi_0$ is the initial porosity and $\hat{\Phi}\equiv\Phi-\Phi_0$ is normalised porosity. Equation \eqref{eq. J = 1+U_Z} may therefore be simplified to
\begin{equation}\label{Eq. Uz=Phi)}
    U_Z = \hat{\Phi},
\end{equation}
meaning strain and normalised porosity are equivalent in this model. Continuity for the fluid phase is given by
\begin{equation}\label{eq. fluid cont}
    \frac{d \hat\Phi}{d t} + \frac{\partial Q}{\partial Z} = 0
\end{equation}
where $Q(Z,t)$ is the Lagrangian relative flow (flux for short) and $d/dt$ is the material derivative with respect to the solid (the time derivative for an observer attached to a solid particle).

Assuming Darcy's law to govern the relative flow of the fluid with respect to the solid, neglecting gravity and inertia \footnote{The porous medium Reynolds number is given by $Re=|\tilde Q|d/\rho\mu$ \citep{cheng:2016}, where $|\tilde Q|= L_0/T_{pe}|Q|\sim O(10^{-3}-10^{-4})$ is the flux magnitude (using parameter orders reported in \ref{subsec: params} and from \cite{godard:2025}), $d\sim O(10^{-5}-10^{-3})$ is the pore size, $\rho\sim O(10^3)$ and $\mu\sim O(10^{-3})$ are the density and viscosity of water, giving $Re\sim O(10^{-6}-10^{-8})\ll 1$. The assumption that the acceleration of the fluid can be neglected and hence of the solid skeleton, whose deformation depends entirely on the fluid flow, is therefore justified.}, flux $Q$ is given by
\begin{equation}\label{Eq: Darcy}
    Q = -\frac{k(\hat\Phi)}{J}\frac{\partial P}{\partial Z}.
\end{equation}
with $P(Z,t)$ and $k$ fluid pressure and permeability as before. Permeability is assumed to be porosity dependent, and the normalised Kozeny-Carman law widely used for a range of materials is adopted, as given in \citet{macminn:2016}:
\begin{equation}\label{eq. permeability}
    k(\hat\Phi) = \frac{(\hat\Phi/\Phi_0+1)^3}{1+\hat\Phi}.
\end{equation}

Let $s(Z,t)$ denote the total nominal stress, which describes the current total force supported by the fluid and solid per reference unit total area. It may be written in terms of the effective stress $s'(Z,t)$ and the fluid pressure $P(Z,t)$:
\begin{equation}
    s = s'-P
\end{equation}
To satisfy mechanical equilibrium, the solid skeleton and the fluid must jointly support the mechanical stress. In the absence of body forces and inertia, this requires
\begin{equation}\label{Eq.MechEq}
    \frac{\partial s'}{\partial Z} = \frac{\partial P}{\partial Z}.
\end{equation}

\subsection{Boundary and initial conditions}

The tendon starts in the reference configuration, corresponding to a relaxed or undeformed state, with initial uniform porosity $\Phi_0$:
\begin{equation}\label{Eq. IC}
    U(Z,0) = 0 \quad\text{and}\quad \hat{\Phi}(Z,t) = 0.
\end{equation}
The bone boundary $Z=1$ is assumed to be fixed and impermeable to fluid, so that:
\begin{equation}\label{Eq. BC Z=1}
   U(Z=1,t)=0,\quad Q(Z=1,t)=0.
\end{equation}
At the muscle boundary $Z=0$ the tendon is assumed to be fully permeable and held at a fixed pressure $P_A$:
\begin{equation}
    P(Z=0,t) = 0.
\end{equation}
In addition, the muscle boundary is subjected to a constant tensile stress $s^*$ (loading) or is left free (unloading):
\begin{equation}
    s'(0,t) = s^* \quad\text{or}\quad s'(0,t) = 0.
\end{equation}

\subsection{Summary of non-dimensional model}\label{subsec. summary}

Equations \eqref{eq. Jacphi},\eqref{eq. fluid cont},\eqref{Eq: Darcy} and \eqref{Eq.MechEq} combine to give a non-linear diffusion equation for (normalised) porosity:
\begin{equation}\label{eq. diffusion eq porosity}
    \frac{d\hat{\Phi}}{dt} = \frac{\partial}{\partial Z}\left(D(\hat{\Phi},s')\frac{\partial \hat{\Phi}}{\partial Z}\right)
\end{equation}
where $D$ is the diffusion coefficient given by
\begin{equation}\label{eq: diff coeff}
    D = \frac{(\hat{\Phi}/\Phi_0+1)^3}{(1+\hat{\Phi})^2}\frac{\partial s'}{\partial \hat{\Phi}}
\end{equation}
with two choices of stress-law: the NH law $s'_{NH}$ given by
\begin{equation}\label{eq. neo law}
    s'_{NH}(Z,t) = 2\frac{C_1^m}{E_f}(J-\frac{1}{J})+2\frac{D_1^m}{E_f}(J-1),
\end{equation}
and the FIB law $s'_{FIB}$ given by 
\begin{equation}\label{eq. uncrimp law}
    s'_{FIB}(Z,t) = c_fs_f(Z,t)+(1-c_f)s_m(Z,t),
\end{equation}
where $J=J(Z,t)=1+\hat{\Phi}(Z,t)$. In the FIB law, the effective stress $s_m$ and strain $e_m$ of the NCM,
\begin{equation}\label{eq. NCM components}
    s_m = 2\left[\frac{C_1^T}{E_f}(J-\frac{1}{J})+\frac{D_1^T}{E_f}(J-1)\right] \quad\text{and}\quad e_m = U_Z=\hat{\Phi}
\end{equation}
are separated from from the effective stress $s_f$ and strain $e_f$ of the fibrils,
\begin{equation}\label{eq. fibril components}
    s_f = e_f = \begin{cases}
        \frac{1}{3\sin^2\theta_0}(2J-3+J^{-2}) \quad\text{for}\quad 1\leq J<\sec\theta_0 \\
        \frac{2(1-\cos^3\theta_0)}{3\sin^2\theta_0}J-1 \quad\text{for}\quad J>\sec\theta_0
        \end{cases}.
\end{equation}

Equations \eqref{eq. diffusion eq porosity}-\eqref{eq. fibril components} are solved numerically for $\hat\Phi$ subject to the boundary conditions $\left[D(\hat{\Phi})\partial \hat{\Phi}/\partial Z\right]_{(1,t)} = 0$ and $\hat{\Phi}(0,t)=\Phi^*$, where $\Phi^*$ is calculated by inverting the stress law given $s^*$, and initial condition $\hat{\Phi}(Z,0)=0$, using a finite volume method in space and an implicit variable-step size method in time via the function ode15s in MATLAB (see \citet{godard:2025} for details). Note that the numerical solver for the FIB model is significantly slower than for the NH model, particularly during the fibril recruitment stage, despite remaining in a one-dimensional framework (though this is still in the order of a few minutes).

Stress $s'$ is directly computed using \eqref{eq. neo law}-\eqref{eq. fibril components}. Pressure $P$ is then calculated by integrating $\partial s'/\partial Z$, flux $Q$ using \eqref{Eq: Darcy} and displacement $U$ is computed by numerically integrating $\hat{\Phi}$:
\begin{equation}\label{eq. disp and pressure}
    P(Z,t) = s'(Z,t)-s^*; \quad Q(Z,t) = -\frac{k}{J}\frac{\partial s'}{\partial Z}; \quad\ U(Z,t) = -\int_1^Z\hat{\Phi}(Z',t)dZ'.
\end{equation}

In addition, let the average stress and average strain be respectively given by:
\begin{equation}\label{eq: average stress and strain}
    \overline{s}(t) := \int_0^{1} s'(Z,t)dZ; \quad \overline{e}(t) :=  \int_0^1 U_Z(Z,t)dZ = -U(0,t).
\end{equation}

\subsection{Parameters of the models}\label{subsec: params}

The choice of parameters will play an important role in comparing the NH and FIB models, namely: the Young's moduli ($E_f$, $E_m$, $E_T$), the maximum crimp angle $\theta_0$, the fibril fraction $c_f$ and Poisson's ratio $\nu$.

Ranges for these parameters are reported in table \ref{tab:params} from various experiments on human tendons, mainly on the patellar tendon. Crimp angles (in radians) are reported from experiments on Achilles tendon, and initial porosity is reported for a range of tendons. A dash is left where the parameter value is unknown. 

\begin{table}[h]
\centering
\renewcommand{\arraystretch}{1.5}
\begin{tabular}{ c c c c c}
 \textbf{Parameter} & \textbf{Symbol} & \textbf{Range or value} & \textbf{Tendon} &\textbf{Reference(s)} \\
 \hline
 Fibril Young's modulus & $E_f$ & $2.8\pm0.3 \text{GPa}$ & Patellar & \citep{svensson:2011} \\  
 \hline
 NCM Young's modulus & $E_m$ & - & - &  \\
 \hline
 Tendon Young's modulus & $E_T$ & $0.6-2.2 \text{GPa}$ & Patellar & \citep{bojsen-moller:2007,couppe:2009,onambele:2007,rees:2002,helland:2013,kongsgaard:2010,svensson:2011}\\
 \hline
 Maximum crimp angle & $\theta_0$ & $0.23 - 0.36$ & Achilles & \citep{jarvinen:2004} \\
 \hline
 Fibril fraction & $c_f$ & $0.64-0.85$ & Patellar & \citep{svensson:2011,goh:2008}\\
 \hline
 Initial porosity & $\Phi_0$ & 0.55-0.7 & Various & \citep{thorpe:2015}
\end{tabular}
\caption{Parameter symbols and range of reported values for human tendons}
    \label{tab:params}
\end{table}

The fibril Young's modulus is greater than the whole tendon one, reflecting that fibrils are stiff but do not make up the entire tendon. Although there are no reported values for $E_m$, we expect the NCM to be much softer than the fibrils since they are the main load-bearing unit of tendon; for example we could set $E_m=0.01E_f$. One way of consistently prescribing the whole tendon modulus is to set $E_T = c_fE_f + (1-c_f)E_m$. Since $E_m\ll E_f$ then $E_T \approx c_fE_f$. 

The choice of Poisson's ratio for a poroelastic tendon model is not obvious. In poroelasticity there are two relevant Poisson's ratios: the undrained ratio, measured under  ``undrained'' conditions (i.e. the fluid is not allowed to leave the material), and the drained ratio measured under ``drained'' conditions (fluid moves freely) \citep{cheng:2016}. In our model the tendon will be loaded sufficiently slowly to allow fluid flow, so that the neo-Hookean law (in both the FIB and NH models) requires the drained Poisson's ratio, which is known to be smaller than the undrained ratio. In experiments measuring tendon's Poisson ratio however the tissue is typically assumed to be purely elastic (or viscoelastic), so that fluid flow is not accounted for. This may explain the large variability in reported ratios \citep{gatt:2015,vergari:2011,thorpe:2014}, from negative values to values greater than $1/2$. Due to the lack of agreement in the literature and the idealised nature of our model, we consider values of $\nu$ in the typical range $0\leq\nu<1/2$, and compare the resulting stress law for $\nu=0$ and $\nu=0.3$. 

\begin{figure}[h!]
\centering
   \subfloat[$\nu=0$ \label{subfig: nu=0}]{%
      \includegraphics[width=0.99\textwidth]{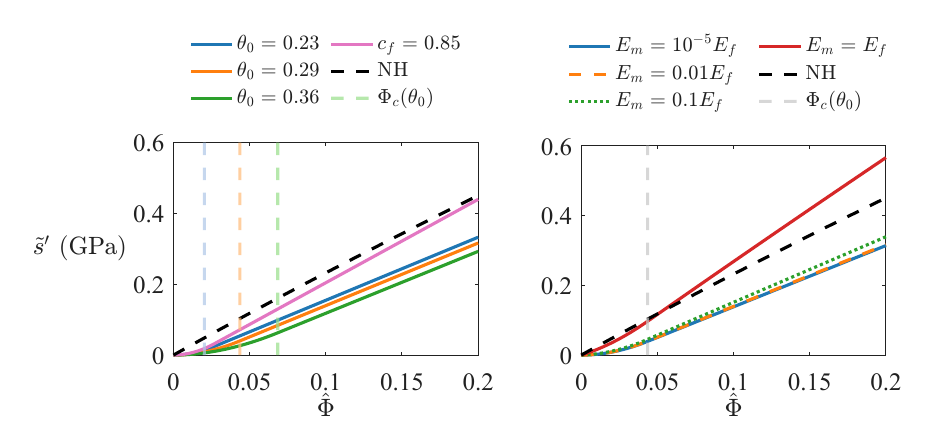}
    }
    \hfill
    \subfloat[$\nu=0.3$ \label{subfig-2: nu=0,3}]{%
      \includegraphics[width=0.99\textwidth]{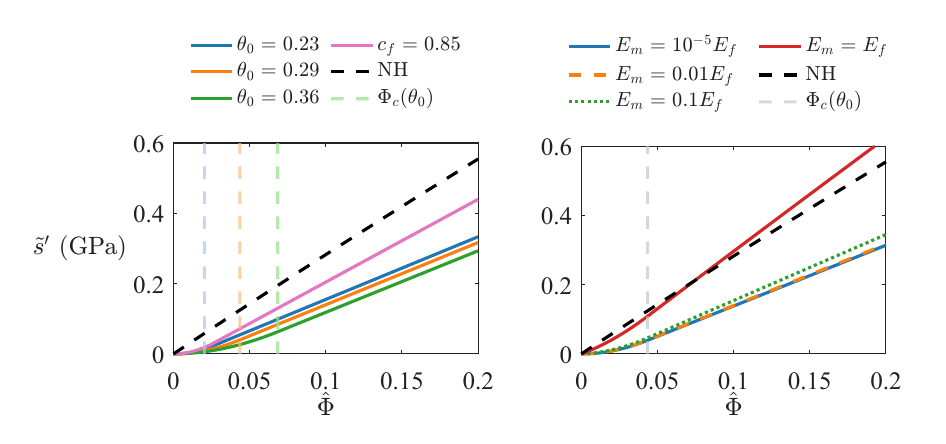}
    }
     \caption{Effective stress against strain / normalised porosity with fixed $E_m=0.01E_f$, varying $\theta_0$ (left) and fixed $\theta_0=0.29$, varying $E_m$ (right) for two values of $\nu$. Fibril fraction is fixed at $c_f=0.64$ in all cases except one where $c_f=0.85$ (pink line). The threshold $\hat{\Phi}_c(\theta_0)$ for each $\theta_0$ is plotted as a dashed line in a lighter shade of the corresponding colour. The neo-Hookean stress law is also plotted in dashed black. The FIB stress law is softer than the NH law for most parameter values.}
     \label{fig:stress_laws_comparison}
\end{figure}

In Figure \ref{fig:stress_laws_comparison} we plot dimensional NH and fibril stresses $\tilde{s}'$ as a function of strain/normalised porosity (recalling that $U_Z=\hat{\Phi}$), setting $E_f=2.8\text{GPa}$ and $c_f=0.64$, so that $E_T\approx 1.8\text{GPa}$. These specific values are chosen from \citet{svensson:2011} who performed mechanical tests on patellar tendons \textit{in vivo} and on fibrils from the same tendons \textit{ex vivo}. This is plotted for the two chosen values of $\nu$.

On the left hand side, we fix the NCM modulus to $E_m = 0.01E_f$ and vary $\theta_0$ with fixed $c_f=0.64$. The intermediate value $\theta_0=0.29$ is the average reported crimp angle for healthy Achilles tendon in \citet{jarvinen:2004} (experiments were not performed on patellar tendon). In addition, in pink we plot the FIB stress for fixed $\theta_0=0.23$ and $c_f=0.85$ (but maintaining $E_T\approx1.8\text{GPa}$). The threshold $\hat{\Phi}_c(\theta_0) = J_c(\theta_0)-1$ beyond which all fibrils are taut is plotted for each $\theta_0$ in dashed horizontal lines. 

A larger maximum crimp $\theta_0$ results in a longer toe region (i.e. the toe region spans a greater interval of $\hat{\Phi}$), resulting in a softer tendon. Note the increase in toe region with $\theta_0$ is significant: the $57\%$ increase in $\theta_0$ from $0.23$ to $0.36$ results in a $153\%$ increase in $\hat\Phi_c(\theta_0)$. As expected for $c_f=0.85$, the upper end of the fibril fraction range, the FIB stress is larger than for $c_f=0.64$, however, within the physiological range of strain it remains smaller than the NH stress (note this is even with $\theta_0=0.23$). In addition, this is comparing to the NH stress where $E_T$ is calculated using $c_f=0.64$; if we had computed it with $c_f=0.85$ then $E_T\approx 2.4\text{GPa}$ and the resulting NH stress would be even higher in comparison.

On the right hand side, we fix $\theta_0=0.29$ and $c_f=0.64$ and now vary the NCM modulus $E_m$ (maintaining $E_T\approx1.8\text{GPa}$). We include $E_m=E_f$ for completeness, though as noted for tendon it is reasonable to assume $E_m\ll E_f$. The plot reveals there is little quantitative difference between taking $E_m=10^{-5}E_f$ and $E_m=0.1E_f$.

These observations are valid for both $\nu=0$ and $\nu=0.3$, and there is very little qualitative difference between the two. Increasing $\nu$ mostly affects the NH model quantitatively, producing a stiffer law, however it makes very little quantitative difference to the FIB stress law as the neo-Hookean background matrix is so soft. Since in \citet{godard:2025} we set $\nu=0.3$, we choose $\nu=0.3$ for this study, but note that the qualitative comparison between the NH and FIB models remains valid for any $0\leq\nu<1/2$.

For our various possible parameter choices within the tendon range, the FIB stress law is softer than the NH law despite the fibril Young's modulus being greater than the tendon Young's modulus used in the NH law. This is due to the very soft background matrix in the FIB law which constitutes a non-negligible fraction of the material.

Based on the observations above, we choose the following base parameters for our model:
\begin{equation}
    \begin{gathered}
      c_f = 0.64 \quad \theta_0 = 0.29 \quad \nu = 0.3 \quad \Phi_0=0.55 \\
      E_f = 2.8\text{GPa} \quad E_m = 0.01E_f \quad E_T = c_fE_f+(1-c_f)E_m\approx1.8\text{GPa}.
\end{gathered}
\end{equation}
For the non-dimensional model, the only dependence on initial porosity is via permeability \eqref{eq. permeability} which increases with $\Phi_0$. Within the reported range $0.55-0.7$, the value of $\Phi_0$ has little quantitative and no qualitative impact on permeability; the choice of $\Phi_0=0.55$ is arbitrary and qualitative results remain unchanged for other values in the range. The value of $\Phi_0$ would be relevant if the exact water content were required ($\Phi$ as opposed to $\hat\Phi$).

There remains to choose a value for $s^*$. The maximum reported strain for the whole tendon in \citet{svensson:2011} is $\Phi^* = 0.06$. Considering this value in the FIB law yields $\tilde{s}_{FIB}^*\approx0.069\text{GPa}$ (slightly higher than the reported maximum stress of $0.054\text{GPa}$ from \citep{svensson:2011}). We choose $\tilde{s}^*=0.069\text{GPa}$ so as to remain in the reported strain range in both models (since the FIB model is softer than the NH model).

\section{Results}\label{sec: results}

We now investigate the tendon's response to loading and unloading at the muscle boundary. In \ref{subsec: loading} the tendon is statically loaded until it reaches steady state and the FIB and NH models are compared. The fibril and NCM components of the FIB model response are then separated and analysed. In \ref{subsec: load-unload}, a constant static load is applied at the muscle boundary until steady state is reached and then released. Again, the FIB and NH models are compared and analysed.

\subsection{Loading}\label{subsec: loading}

We begin by loading the tendon with a constant static stress $s^*$ at the muscle boundary ($Z=0$) until steady state is reached:
\begin{equation}
    s^* = 0.069\text{GPa}/E_f \approx 0.025,
\end{equation}
which is the stress calculated in Section \ref{subsec: params} using the reported maximum strain from \citet{svensson:2011}. For this applied stress the corresponding ``applied strain'' for the FIB model falls in the regime $\Phi^*_{FIB}>\hat{\Phi}_c(\theta_0)$ so that the tendon may operate in the linear regime where all fibrils are taut.

\subsubsection{Model comparison}

\begin{figure}[h!]
    \centering
    \includegraphics[width=\linewidth]{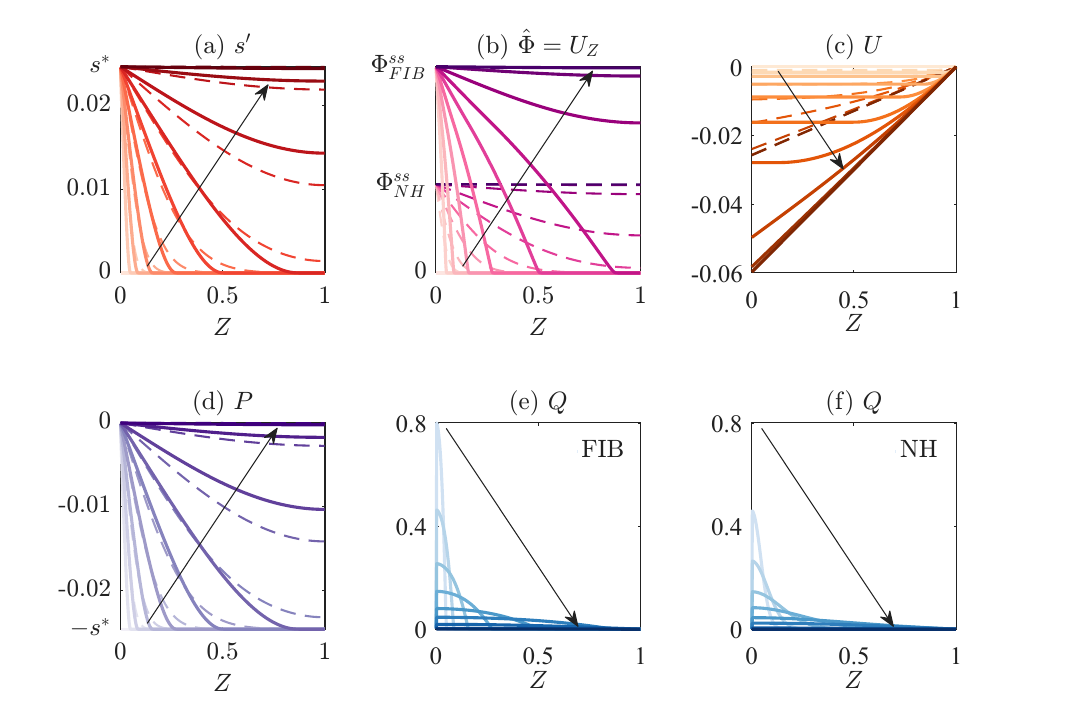}
    \caption{Solid stress (a), solid strain or porosity (b), solid displacement (c), fluid pressure (d) and fluid flux (e-f) against $Z$ for $t = 0,0.001,0.003,0.01,0.03,0.1,0.3,1,2,3$ increasing from light to dark (arrow is increasing time). Solid lines are for the FIB model and dashed lines are for the NH model, except for flux which is separated into two plots for clarity. All quantities reach steady state within the simulation time. Note $\Phi^{ss}_{FIB}\approx 0.06$ and $\Phi^{ss}_{NH}\approx 0.029$.}
    \label{fig:full_static}
\end{figure}

In Figure \ref{fig:full_static}, we plot solid stress (a), porosity or equivalently solid strain (b), solid displacement (c), fluid pressure (d) and fluid flux (e-f) against $Z$ at several time points, for the FIB model (solid lines) and the NH model (dashed lines, except for flux which is separated into two plots for clarity). All quantities reach\footnote{Note that while formally quantities only tend towards steady state due to the diffusive nature of the system, the difference between our solution and the steady state is negligible; the term ``reach'' is employed here and throughout the paper for simplicity.} steady state within the simulation time. In both the fibril and NH models, tendon stress and strain reach their steady state values $s^{ss}=s^*$ and $\Phi^{ss} = {\Phi}^*$ for all $Z$, pressure and flux relax to zero for all $Z$ and displacement tends to linear in $Z$ over time, where maximum displacement is achieved at $Z=0$. The steady state stress $s^{ss}$, and thus maximum pressure magnitude $P(Z,t=0) = -s^{ss}$ (c.f.\ eq. \eqref{eq. disp and pressure}), is the same in both models (\ref{fig:full_static}a,d). On the other hand, since the NH tendon is stiffer than the FIB tendon, the steady state strain for the FIB model is greater than for the NH model: $\Phi^{ss}_{FIB}\approx 0.06 >\Phi^{ss}_{NH}\approx0.029$ (\ref{fig:full_static}b). Accordingly, displacement throughout the material and maximum flux are greater in the FIB model than the NH model (\ref{fig:full_static}c,e,f). In fact, the maximum flux is almost twice as large and the maximum displacement is almost three times as large in comparison with the NH model. Since the material is stretched more in the FIB model, the incompressibility condition requires a greater inflow to fill the larger pores. In addition, while the NH stress, porosity and flux display smooth profiles in $Z$, decaying smoothly to zero from $Z=0$ to $Z=1$, the FIB stress, porosity and flux profiles exhibit sharp turning points as their values approach zero at a given $Z =: Z_c$. This $Z_c$ corresponds to the position at which fibrils have not yet been recruited, and so where the material is still very soft: $Z_c = Z|_{\hat{\Phi}=\hat{\Phi}_c(\theta_0)}$. In other words, $R_c(Z)=0$ for $Z>Z_c$ and $R_c(Z)>0$ for $Z\leq Z_c$.

\begin{figure}[h!]
\centering
\subfloat[\label{fig:Phi_end_to_steady_state}]{%
      \includegraphics[width=0.48\textwidth]{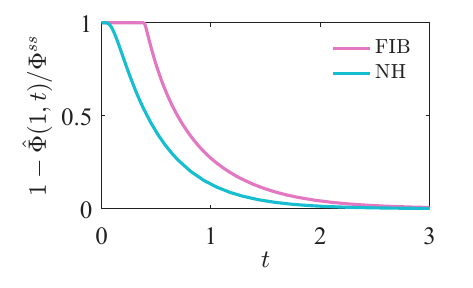}
    }
    \hfill
    \subfloat[\label{fig:ds_dPhi_vs_Phi}]{%
      \includegraphics[width=0.48\textwidth]{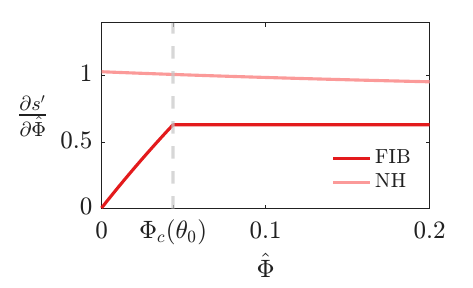}
    }
     \caption{(a) Evolution of $\hat{\Phi}(1,t)$ to steady state $\Phi^{ss}$. The NH porosity reaches steady state faster than the FIB porosity. (b) Rate of change of stress with porosity (strain) against porosity; $\partial s'/\partial \hat{\Phi}$ for the NH law is greater than for the FIB law, even when all fibrils are uncrimped.  }
\end{figure}

The NH and FIB models also approach their steady states at different rates. Defining relative porosity as $\hat\Phi/\Phi^{ss}$, this is clearly seen in Figure \ref{fig:Phi_end_to_steady_state} which for each model displays the evolution of porosity at $Z=1$ to their respective steady states $\Phi^{ss}_{FIB}$ and $\Phi^{ss}_{NH}$: the NH porosity approaches its steady state faster than the FIB porosity. This is due to the diffusive nature of the system (c.f. equation \eqref{eq. diffusion eq porosity}), so that the rate at which porosity reaches its steady state is governed by the diffusion coefficient $D(\hat{\Phi},s')$. This means diffusion is greater in the NH model than in the FIB model: $D_{NH}>D_{FIB}$.

Diffusion (given by \eqref{eq: diff coeff}) depends on porosity, permeability and the rate of change of stress with porosity (strain) $\partial s'/\partial\hat{\Phi}$, which is a local measure of the tissue's stiffness. Note this also means diffusion depends on $Z$. The difference between $D_{FIB}$ and $D_{NH}$ here is mainly governed by the difference in $\partial s'/\partial \hat{\Phi}$, which is plotted in Figure \ref{fig:ds_dPhi_vs_Phi}. In particular, the FIB $\partial s'/\partial \hat{\Phi}$ is very close to zero for $\hat{\Phi}=0$ so that initially diffusion is also close to zero, however it then increases rapidly with $\hat{\Phi}$ as fibrils are recruited ($0<\hat{\Phi}<\hat{\Phi}_c(\theta_0)$) and is constant once all fibrils have been recruited ($\hat{\Phi}\geq\hat{\Phi}_c(\theta_0)$). This is in contrast to the NH model for which $\partial s'/\partial \hat{\Phi}$ is finite for $\hat{\Phi}=0$ and gradually decreases with porosity. Nevertheless, $\partial s'/\partial \hat{\Phi}$ for the FIB model remains smaller than in the NH model, resulting in a smaller diffusion. The sharp corners observed in the FIB porosity/strain and flux in Figure \ref{fig:full_static} can also be explained by the rapid increase in diffusion from near zero at $Z=Z_c$.

We have identified three key differences between the NH and FIB models: the NH model approaches steady state faster than the FIB model, the FIB model is stretched more than the NH model, and the FIB model exhibits sharp corners in its stress, porosity and flux profiles in $Z$ while the NH model does not. The first two points are simply a result of the NH model being stiffer than the FIB, due to the soft background matrix of the FIB model. As discussed however, even increasing the fibril fraction to $c_f=0.85$ and decreasing the crimp angle to $\theta_0=0.23$ would maintain the FIB model as softer. The only scenario in which the FIB model may be stiffer than the NH model is if the background matrix were also stiff, which is not the case for tendons. The third point is a consequence of the fibril recruitment rule, dictated by the crimp angle distribution $\theta(r)$, and is qualitatively independent of the choice of parameters. Here we chose the distribution to depend only on a maximum crimp angle $\theta_0$, which determines where the sharp corners appear.

\subsubsection{FIB model: NCM and fibril response during loading}

\begin{figure}[h!]
    \centering
    \subfloat[$s^*\approx0.025$ \label{subfig-1: s*=0.025}]{\includegraphics[width=0.95\textwidth]{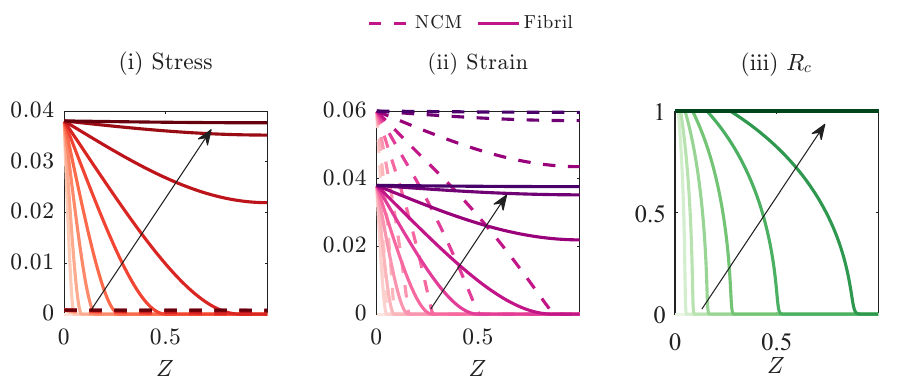}}
    \hfill
    \subfloat[$s^*\approx0.001$ \label{subfig-2: s*=0.01}]{\includegraphics[width=0.95\textwidth]{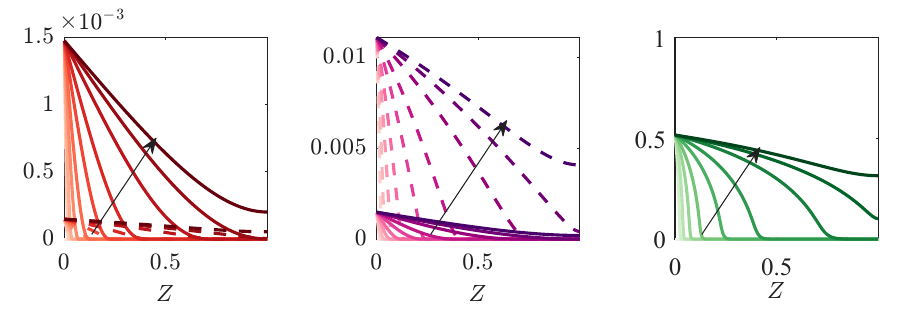}}
    \caption{NCM (dashed) and fibril (solid) stress (i) and strain (ii) and fraction of taut fibrils $R_c$ (iii)  against $Z$ for $t = 0,0.001,0.003,0.01,0.03,0.1,0.3,1,2,3$ increasing from light to dark (arrow is increasing time), for two applied loads. The fibrils bear almost all the load, which increases as $R_c$ increases (fibrils uncrimp). The strain however is shared between the fibrils and the NCM; the NCM takes on a large portion of the strain when fibrils are crimped, which decreases as fibrils are recruited.}
    \label{fig:fibril_NCM_stress_strain}
\end{figure}

We now restrict our attention to the FIB model only. In addition to applying a load of $0.025$, we also apply a smaller load of $0.01$, so that the corresponding ``applied strain'' is approximately $0.01\ll\hat\Phi_c(\theta_0)$ (i.e.\ not all fibrils will be taut). In Figure \ref{fig:fibril_NCM_stress_strain} we plot the NCM and fibril components of the stress (i) and strain (ii) as defined in \eqref{eq. NCM components} and \eqref{eq. fibril components} for the two applied loads. Also plotted is the non-dimensional radius of taut fibrils $R_c$ (iii), calculated at each $Z$ and $t$ using equation \eqref{eq. Rc}. Here, this simply represents the fraction of taut fibrils at a given $Z$ and $t$: for $R_c=0$ all fibrils there are crimped and for $R_c=1$ all fibrils there are taut.

In both cases, as the fibrils uncrimp ($R_c$ increases), the load borne by the fibrils increases ($s_f$ increases) and the fibril and NCM strain increase ($e_f$ and $e_m$ increase), while the NCM stress remains close to zero ($s_m\approx0$). For $s^*\approx0.025$ (\ref{subfig-1: s*=0.025}), all fibrils uncrimp ($R_c$ reaches 1 everywhere), allowing the fibrils to bear a considerable fraction of the strain, and practically all the stress. The steady state NCM strain is larger than the steady state fibril strain. Note that it takes approximately $t=1$ for all the fibrils to become taut for all $Z$ however it takes until approximately $t=3$ for the porosity to reach steady state. For $s^*=0.01$ (\ref{subfig-2: s*=0.01}) however, not all fibrils uncrimp ($\max_{R_c}\approx0.5$) and the system does not reach a steady state within the simulation time. The fibrils still bear most of the load, though a smaller proportion than in the former case, but bear a much smaller fraction of the strain, which is mostly taken on by the NCM. 

The fibrils bear most of the load since they are 100 times stiffer than the NCM. In parallel, the fibrils' high stiffness and uncrimping feature protect them from excessive strains, while the soft NCM takes on a larger proportion of the strain. At small loads, where not all fibrils are uncrimped, the NCM bears most of the strain and some of the stress which spares the fibrils from unnecessary strain. At larger loads, most or all fibrils are recruited and provide the high tensile strength which prevents the tendon and NCM from overstraining. The recruitment process therefore protects fibrils, and their high stiffness protects both fibrils and NCM, from overstraining. This provides tendon with a natural protection mechanism against large loads, without compromising its ability to stretch, key to its mechanical function.

\subsection{Unloading}\label{subsec: load-unload}

In order to examine the impact of fibres in the relaxation phase of a load/unload cycle, we now subject the material to piecewise constant stress at the boundary, such that:
\begin{equation}
    s'(0,t) = \begin{cases}
        s^* \quad&\text{for}\quad 0\leq t\leq t_1 \\
        0 \quad&\text{for}\quad t_1<t\leq t_2,
    \end{cases}
\end{equation}
with $s^*=0.069\text{GPa}/E_f\approx0.025$ as before.

\begin{figure}[h!]
\centering
\subfloat[\label{fig:load-unload-cycle}]{%
      \includegraphics[width=0.99\textwidth]{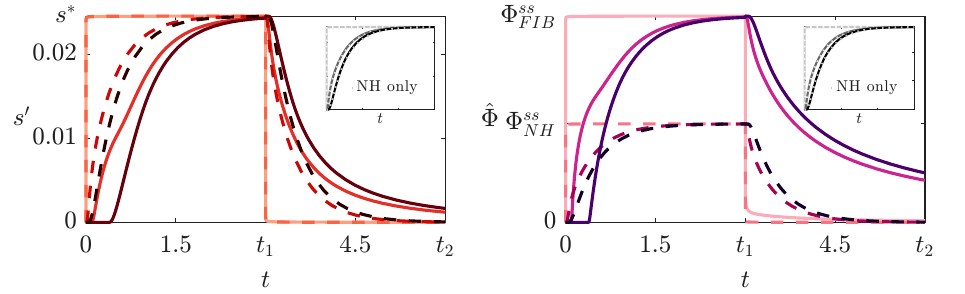}
    }
    \hfill
\subfloat[\label{fig:load-unload-Phi-Rc-to-steady-state}]{%
      \includegraphics[width=0.99\textwidth]{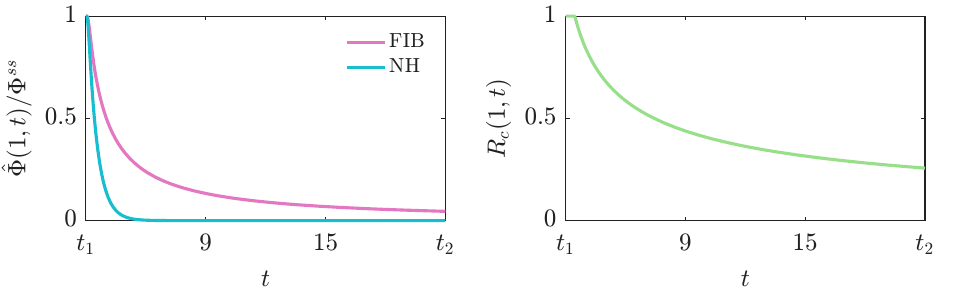}
    }
     \caption{(a) Stress (left) and porosity (right) at $Z={0,0.5,1}$ (increasing from light to dark) against time during loading and unloading $t_2=2t_1=6$ for the FIB model (solid) and NH model (dashed). The insets display NH $s'(t)$ and $\hat\Phi(t)$ (loading, solid lines) and $s^*-s'(t-t_1)$ and $\Phi^{ss}_{NH}-\hat\Phi(t-t_1)$ (unloading, dotted lines) at $Z={0,0.5,1}$, illustrating overlap.  Stress and porosity reach steady state during loading and unloading in the NH model, but not during unloading in the FIB model. (b) Evolution of porosity (left) and fraction of taut fibril (right) at $Z=1$ to steady state during unloading with $t_2=7t_1=21$. The NH model reaches steady state but the FIB model does not, and there remains a non-zero fraction of taut fibrils.}
\end{figure}

In Figure \ref{fig:load-unload-cycle} we plot stress and porosity against time for three values of $Z = 0,0.5,1$, increasing from light to dark, for both the NH and FIB models, with $t_2 = 2t_1 = 6$. In addition, insets display, in greys, NH $s'(t)$ and $\hat\Phi(t)$ during loading and $s^*-s'(t-t_1)$ and $\Phi^{ss}_{NH}-\hat\Phi(t-t_1)$ during unloading. The FIB stress and porosity (coloured solid lines) exhibit an asymmetry in the loading and unloading phases: the stress and porosity reach the loading steady state ($s'=s^*$ and $\hat{\Phi}=\Phi_{FIB}^{ss}$) throughout $Z$ during the loading phase but do not reach the unloading steady state ($s'=\hat{\Phi}=0$) for all $Z$ by the end of the unloading phase ($t=6$). The NH stress and porosity on the other hand are visually symmetric in the loading and unloading phase, as highlighted by the matching loading and unloading curves shown in the insets, and stress and porosity reach steady state for all $Z$ in both phases ($s^*$, $\Phi^{ss}_{NH}$ in loading and zero in unloading).

This slow relaxation for the FIB model during unloading is highlighted in Figure \ref{fig:load-unload-Phi-Rc-to-steady-state} which displays the evolution of relative porosity (i.e.\ scaled by $\Phi^{ss}$) and the fraction of taut fibrils $R_c$ at $Z=1$ against time for the unloading phase only, now setting $t_2 = 7t_1=21$ (i.e. $t_1=3$ and we let the material relax for a time of $t=18=6t_1$). Although $\hat{\Phi}(1,t)$ approaches zero it remains finite even at $t=21$. As a consequence, there remains a considerable fraction of taut fibrils at the end of the loading phase ($R_c(Z=1,t=t_2)\approx 26\%$).

As seen in the previous section, the FIB porosity reaches steady state during loading relatively quickly (by $t=3$) since diffusion increases as fibrils are recruited. During unloading however, as porosity decreases, fibrils become crimped once again and diffusion approaches zero. Indeed, referring back to Figure \ref{fig:ds_dPhi_vs_Phi} we see that $\partial s'/\partial \hat{\Phi}$ which governs diffusion remains constant while $\hat{\Phi}\geq\hat{\Phi}_c(\theta_0)$, however once $\hat{\Phi}<\hat{\Phi}_c(\theta_0)$ (not all fibrils recruited) the gradient rapidly decreases and diffusion approaches zero, slowing the response in $Z$. The smaller porosity is, the smaller $R_c$ is, the less fibrils bear load and the softer the tendon becomes, tending diffusion towards zero. This generates a feedback loop which continues to slow the relaxation in time. With diffusion close to zero, the material approaches its steady state at a much slower rate.

\begin{figure}[h!]
\centering
    \subfloat[\label{fig:total-stress-vs-strain}]{%
      \includegraphics[width=0.99\textwidth]{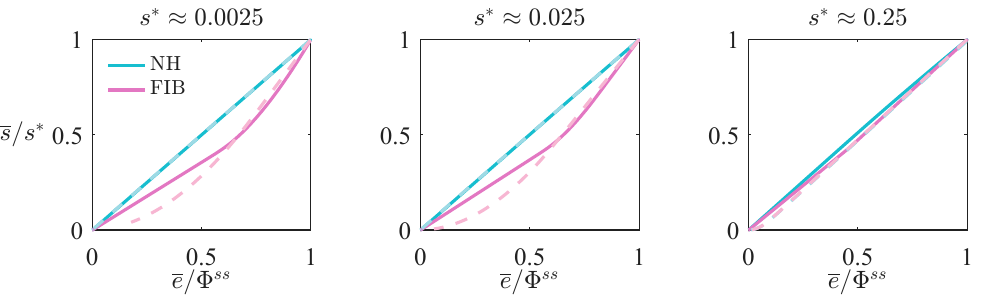}
    }
\hfill
    \subfloat[\label{fig:Rc_heatmap}]{\includegraphics[width=0.99\textwidth]{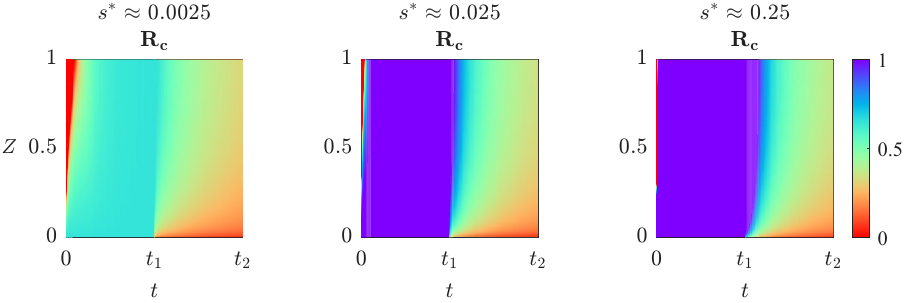}}
\hfill
    \subfloat[\label{fig:load-unload-Phi-Rc-to-steady-state-3As}]{%
      \includegraphics[width=0.99\textwidth]{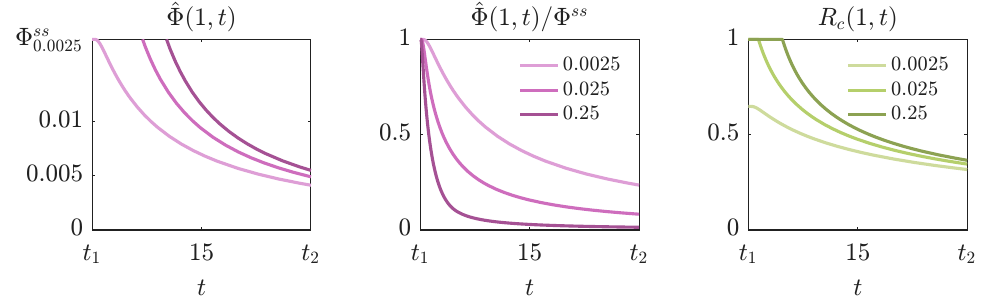}
    }
    
     \caption{(a-c) $t_2=2t_1 = 20$. (a) Normalised average stress against normalised average strain, during loading (solid lines) and unloading (dashed lines), for three values of $s^*$. The loading and unloading curves match in the NH model but not in the FIB model. (b) Heatmap of evolution of $R_c(Z,t)$ for three values of $s^*$. During unloading only fibrils near $Z=0$ are fully recrimped, leaving a non-negligible fraction of taut fibrils in the rest of the tissue. (c) Evolution of FIB porosity (left), zoomed in to $0\leq\hat{\Phi}\leq\Phi^{ss}_{0.0025}$ to distinguish the three lines, where $\Phi^{ss}_{0.0025}$ is the porosity steady state associated with $s^*\approx0.0025$; relative porosity (middle); and fraction of taut fibrils at $Z=1$ during unloading (right), for three values of $s^*$. Relative porosity $\hat{\Phi}/\Phi^{ss}$ relaxes faster with increasing $s^*$, while $\hat{\Phi}$ and $R_c$ increase with increasing $s^*$.}
\end{figure}

In Figure \ref{fig:total-stress-vs-strain} we plot the normalised average stress $\bar{s}/s^*$ against normalised average strain $\bar e/\Phi^{ss}$ for both the NH and FIB models, for three values of $s^*$ (normalisation allows for consistent comparison across the three values). The average stress and strain, defined in \eqref{eq: average stress and strain}, are averages along the length of the tendon. Solid lines and dashed lines represent loading and unloading respectively. Since $s^*\approx 0.0025<s'_{FIB}(\hat{\Phi}_c(\theta_0))=0.014$, not all fibrils will be recruited during loading for this value of $s^*$, meaning the tissue will remain relatively soft and so will take longer to reach steady state; we therefore set $t_2=2t_1=20$ in order for all cases to reach steady state during loading. In the NH model, the loading and unloading curves match. For the FIB model on the other hand the loading and unloading curves do not match, revealing hysteresis: given an average strain $\overline{e}$, the average stress $\overline{s}$ is lower in the unloading phase than in the loading phase. The hysteresis is significantly reduced for larger stress $s^*\approx 0.25$ (i.e.\ the difference between the loading and unloading curves is smaller). In addition, the unloading FIB curve for $s^*\approx 0.0025$ ends significantly before reaching zero, meaning the average strain relative to $\Phi^{ss}$ in the tissue is non-negligible at $t=t_2$.

Figure \ref{fig:Rc_heatmap} displays heatmaps of the fraction of taut fibrils $R_c$ in time and space for the same three values of $s^*$. At the two larger values of applied stress, all fibrils are recruited in the loading phase ($R_c=1$ everywhere at $t=t_1$), while at the smallest $s^*$ (left plot), only about $65\%$ of fibrils are recruited. During unloading however, we see again a strong asymmetry: only fibrils near $Z=0$ are fully re-crimped, even for the larger applied loads. For example, in the right plot, the taut fibril fraction at $Z=1$ only reaches $R_c\approx0.37$ by the end of the simulation. 

In Figure \ref{fig:load-unload-Phi-Rc-to-steady-state-3As} we plot the evolution of FIB porosity, FIB relative porosity (i.e. scaled by $\Phi^{ss}_{FIB}$) and the fraction of taut fibrils $R_c$ at $Z=1$ against time during the unloading phase and for the three values of $s^*$. Both the porosity and the fraction of taut fibrils show an increasing dependence on applied stress, with $\hat{\Phi}(1,t)$ and $R_c(1,t)$ smallest at the smallest value of $s^*$. However, this is largely due to the fact that the material deforms significantly less at this value. When considering the {\it relative} porosity (middle plot), the trend reverses, with the fastest relaxation at the largest $s^*$. 

The hysteresis observed for the FIB model and the difference in taut fibril fraction between the loading and unloading phases highlights the asymmetry observed in Figure \ref{fig:load-unload-cycle}. Indeed, all three cases of $s^*$ reach steady state during loading since their diffusion coefficients are large enough for the response to the applied stress to propagate throughout the material within the imposed time frame. Note that the diffusion coefficient at the end of the loading phase, which is uniform in $Z$, increases with $s^*$. This means that as the unloading phase begins, the case with larger $s^*$ starts with a much higher $D_{FIB}$, which creates a relatively faster initial relaxation. Nonetheless, as the tissue relaxes, fibrils are re-crimped, softening the material and thus decreasing diffusion. The result is that tissue does not fully relax within the imposed unloading time. For smaller values of $s^*$, on the other hand, the diffusion coefficient is lower at the start of the unloading phase, creating a slower relaxation, but the tissue begins closer to its resting state.

\section{Discussion}\label{sec: discussion}

In this study we employed a non-linear fibre-reinforced poroelastic framework to model a tendon, which accounts for the presence of fluid, crimped fibrils and a soft background matrix (FIB model). By directly comparing to a model with an isotropic neo-Hookean (NH) solid skeleton and no fibrils, we have been able to investigate both qualitatively and quantitatively the impact of fibre recruitment in the response to an applied load.

One immediate result was that the presence of stiff crimped fibrils embedded in a soft background matrix (the NCM) gives rise to a softer tissue than the NH model. Despite setting a Young's modulus for the NH model as a combination of the stiff fibril Young's modulus and the soft background matrix one, yielding a smaller modulus overall than the fibril modulus, the neo-Hookean stress law for uniaxial deformation of a poroelastic material is stiffer than the fibril (FIB) stress law. This remained true for a range of parameters reported for tendon. Consequently, we found that the presence of stiff crimped fibrils allows the tendon to stretch significantly without excessively straining the fibrils or the background matrix, providing a natural protection mechanism for the tendon's structural components to load, as has previously been observed experimentally \citep{screen:2004}.

When applying a constant static load at the muscle boundary and releasing it, we revealed a significant asymmetry between the loading and unloading phases for the FIB model. The inclusion of fluid flow through the porous matrix introduces an inherent timescale so that the response to a load imposed at one end is not instantaneous in the rest of the tissue. The rate at which the response propagates through the tissue is governed by a diffusion coefficient which varies with porosity and stiffness. Although our poroelastic model is effectively one-dimensional, it accounts for the presence of many fibrils with a range of ``resting'' (initial) crimp angles, so that the fibrils are not all simultaneously taut at a given position. We showed that while the fraction of taut fibrils $R_c(Z,t)$ reaches a uniform steady state in $Z$ during loading, it does not during unloading, and a non-negligible fraction of fibrils remain taut at the end of the loading phase. This is particularly important for the diffusion coefficient, which is constant once all fibrils are taut but close to zero when fibrils are crimped. The asymmetry in the response between loading and unloading is a result of the dependence of diffusion on fibril recruitment via stiffness. For physiological loads or smaller (in this study $s^*\leq0.025$), fibrils quickly re-crimp as unloading begins, softening the tissue and thus significantly slowing its relaxation. For larger, non-physiological loads such as ones experienced during resistance training (e.g. $s^*=0.25$), fibrils initially remain taut, allowing the tissue to quickly relax initially and thus reducing asymmetry. Relaxation however leads to re-crimping, softening the material and slowing the relaxation of the tissue which did not reach steady state within the imposed unloading time. This also means the underlying crimp configuration, and thus tendon microstructure, is not the same during loading and unloading, which may play a role in the tendon's mechanical function and possibly in how it responds to damage. 

The asymmetry between loading and unloading also manifested as a hysteresis loop in the average stress-strain curve. This is a known phenomenon in tendon mechanical tests, having been observed both \textit{in vivo} \citep{maganaris:2000, peltonen:2013} and \textit{ex vivo} \citep{ciarletta:2008,screen:2004,schwerdt:1980, schatzmann:1998, thorpe:2015a}. Typically, hysteresis is attributed to tendon's viscoelastic properties, fibre sliding and/or rearrangement of its microstructure such as interfibrillar bonds breaking and reforming. Theoretical and experimental poroelastic studies also report hysteresis, arising from various mechanisms, including competition between interfacial and bulk compression, strain hardening of the solid network \citep{xu:2024}, increase in viscous drag from fast cyclic loading \citep{fiori:2023}, wall friction \citep{lutz:2021,desclaux:2026}, among others \citep{macminn:2015,hewitt:2016,guyer:2011,sobac:2011}. To the authors' knowledge, no study attributes hysteresis in tendon to the interaction between the poroelastic diffusive nature of the system and the built-in fibre recruitment.

Our analysis under different order of magnitude loads also provided a clear qualitative picture of the relationship between applied load and hysteresis. Under larger loads, the system was found to relax more quickly in a relative sense, which could be directly attributed to an increase in the effective diffusion coefficient. However, in an absolute sense the system remained further from the fully relaxed state, due to the rapid decrease in diffusion coefficient. The observed hysteresis occurs due to the diffusion's dependence on stiffness via fibril recruitment; in a higher dimensional model, fibrils should still be aligned axially and only recruited via axial stretch, so that we  expect hysteresis in the axial direction to persist, though the exact form would likely depend on transverse position. In addition, in a higher dimensional model radial fibril recruitment could be explicitly accounted for, rather than captured as an effective contribution to stress, so we might expect hysteresis to also occur in the transverse directions. We note that in higher dimensions the transverse and axial deformation and flow are coupled, making the full system highly complex; these are simply speculations which remain to be explored and constitute an interesting future research avenue.

One clear advantage of a model that directly incorporates fibres is that the choice of parameters is less ambiguous than when employing a classical stress law such as neo-Hookean elasticity. A major challenge with the neo-Hookean law is choosing the material constants - in our case, Poisson's ratio $\nu$ and tendon's Young's modulus $E_T$. In particular, relating reported Poisson's ratios from experimental measurements to our uniaxial poroelastic model is not straightforward, suggesting the need for experiments measuring Poisson's ratio within a poroelastic framework. This however is less important for the FIB model; we showed that since the background matrix is so soft, its precise material constants do not have much impact on the resulting stress law. The required parameters for the FIB model are mostly measurable: fibril crimp angle $\theta_0$, fibril (or collagen) fraction $c_f$ and fibril Young's modulus $E_f$. While the FIB stress law employed in this study is bespoke for the application at hand by accounting for crimp and fibril recruitment, other fibril constitutive laws can readily be considered and incorporated into this framework.

A limitation of this model is that stress tends to zero as fibrils are crimped, which is unlikely to be the case for biological tissues like tendon. This could be addressed by allowing fibrils to bear a small amount of load when crimped, or including additional components which bear load. Another major mechanism not included in this model is fibre-sliding, which plays an important role in extension beyond the toe region \citep{screen:2004}. In this model, fibrils are the only collagenous component; in reality, tendons have a more complex hierarchical collagen structure. A simple extension to this work is to model tendon as composed of fascicles embedded in NCM, with each individual fascicle composed of crimped fibrils embedded in NCM. This would also be more consistent with the fibril crimp distribution observation which was made for tendon fascicles, and not the whole tendon itself \citep{kastelic:1978}.

Tendons primarily bear axial load, supported by their highly aligned collagen structure, so that deformation is dominantly uniaxial. %In the model, we implicitly assumed no transverse deformation by effectively ``fixing'' the transverse boundaries to a slippery wall; this is highly idealised and would be challenging to experimentally. 
A more accurate model is to have a stress-free transverse boundary condition, and either resolve transverse deformation or exploit the small aspect ratio so that transverse deformation is uniform to first order; future work could be directed towards constructing a reduced three-dimensional model which takes into account these effects. Finally, the imposed stress boundary condition at the muscle-tendon interface is an idealised loading condition, motivated by \textit{ex vivo} mechanical tests; this could be compared to a model where displacement is imposed instead of stress and investigate the impact on hysteresis. Within the body, tendons experience a complex loading pattern from the muscle which is also affected by physical and geometrical constraints from surrounding tissues, likely resulting in a combination of imposed stress and displacement. Future work could integrate a more biologically accurate representation of the muscle-tendon boundary.

\section{Conclusion}\label{sec: conclusion}

This work presented a fibre-reinforced poroelastic model which incorporated a fibre-recruitment rule based on fibril crimp, with the aim of elucidating the role of crimped fibrils in tendon mechanics, and more generally in non-linear poroelastic materials. The presence of stiff, crimped fibrils was found to increase bulk extension while protecting fibrils and the NCM from high strains, and introduced asymmetry and hysteresis in the load-unload sequence. The asymmetry and hysteresis are a consequence of the diffusive nature of the system which introduces an inherent timescale in the tissue's response to loading and unloading. The diffusion itself depends on the stiffness of the tissue via fibril recruitment, so that during unloading the tissue softens and relaxation is slowed down in comparison to its faster approach to loading steady state. This means the underlying crimp configuration and thus tendon microstructure is not the same during loading and unloading. To the authors' knowledge, the mechanism leading to hysteresis in this model is novel within both the poroelastic and tendon framework.

The model provides us with access to quantitative measures of spatial and temporal stress and strain fields of the whole tendon, of the non-collagenous matrix and the fibrils, as well as fluid flux through the solid matrix, allowing for comparison to experimental data to validate and further improve the model. The straightforward inclusion of fibrils in the model represents tendons more closely than the neo-Hookean model. In addition, the FIB stress law is more interpretable within the biological context and makes parameter choices more natural. In our previous study, the primary aim was to investigate heterogeneity as a representation of tendon damage. This fibre-reinforced framework allows us to explore damage more concretely: ongoing work includes exploring the impact of ``breaking'' certain fibrils in the solid stress law, as well as extension to two-dimensions.

\section*{Author contributions}
Designed research: all authors;
Conducted research: Z.C.G.;
Wrote paper: all authors
%We strongly encourage authors to include author contributions and recommend using \href{https://casrai.org/credit/}{CRediT} for standardised contribution descriptions. Please refer to our general \href{https://www.rsc.org/journals-books-databases/journal-authors-reviewers/author-responsibilities/}{author guidelines} for more information about authorship.

\section*{Conflicts of interest}
There are no conflicts to declare.

\section*{Data availability}
The code required to reproduce model output is available in the public repository: 
github.com\slash zoegdr\slash crimpore1D.

\section*{Acknowledgments}
The authors gratefully acknowledge the anonymous reviewers for their valuable suggestions. 

\section*{Funding}
Z.C.G. gratefully acknowledges funding from the
Paul Shreder scholarship. For the purpose of open access, the author has applied a CC BY public copyright licence to any author accepted manuscript arising from this submission.

\bibliographystyle{abbrvnat}
\bibliography{refs}

@article{svensson:2011,
author = {Svensson, Ren{\'e} B. and Hansen, Philip and Hassenkam, Tue and Haraldsson, Bjarki T. and Aagaard, Per and Kovanen, Vuokko and Krogsgaard, Michael and Kjaer, Michael and Magnusson, S. Peter},
title = {Mechanical properties of human patellar tendon at the hierarchical levels of tendon and fibril},
journal = {Journal of Applied Physiology},
volume = {112},
number = {3},
pages = {419-426},
year = {2012},
doi = {10.1152/japplphysiol.01172.2011},
    note ={PMID: 22114175},
URL = {https://journals.physiology.org/doi/abs/10.1152/japplphysiol.01172.2011},
eprint = {https://journals.physiology.org/doi/pdf/10.1152/japplphysiol.01172.2011}
}

@article{gupta:2010a,
    author = {Gupta, S and Lin, Jeremy and Ashby, Paul and Pruitt, Lisa},
    title = {A fiber reinforced poroelastic model of nanoindentation of porcine costal cartilage: A combined experimental and finite element approach},
    journal = {Journal of the Mechanical Behavior of Biomedical Materials},
    year = {2009},
    volume = {2},
    number = {4},
    DOI = {10.1016/j.jmbbm.2008.09.003.},
    pages = {326-338}
}

@article{chen:2020, title={Multiscale modelling and homogenisation of fibre-reinforced hydrogels for tissue engineering}, volume={31}, DOI={10.1017/S0956792518000657}, number={1}, journal={European Journal of Applied Mathematics}, author={Chen, M. J. and Kimpton, L. S. and Whiteley, J. P. and Castilho, M. and Malda, J. and Please, C. P. and Waters, S. L. and Byrne, H. M.}, year={2020}, pages={143–171}}

@article{ahmadi:2025,
    author = {Ahmadi, Masoud and McBride, Andrew and Steinmann, Paul and Saxena, Prashant},
    title = {Plane stress finite element modelling of arbitrary compressible hyperelastic materials},
    journal = {Acta Mechanica},
    year = {2025},
    doi = {10.1007/s00707-025-04310-z},
    number = {7},
    volume = {236}
}

@book{ogden:2013,
    author = {Ogden, R.W.},
    title = {Non-Linear Elastic Deformations},
    publisher = {Dover Publications},
    year = {2013}
}

@article{sobac:2011,
  title={Dynamique de mousses poro{\'e}lastiques},
  author={Sobac, Benjamin and Colombani, Mathieu and Forterre, Yo{\"e}l},
  journal={Mechanics \& Industry},
  volume={12},
  number={3},
  pages={231--238},
  year={2011},
  publisher={EDP Sciences}
}

@article{guyer:2011,
  title={Hysteresis in modeling of poroelastic systems: Quasistatic equilibrium},
  author={Guyer, RA and Kim, H Alicia and Derome, Dominique and Carmeliet, Jan and TenCate, J},
  journal={Physical Review E—Statistical, Nonlinear, and Soft Matter Physics},
  volume={83},
  number={6},
  pages={061408},
  year={2011},
  publisher={APS}
}

@article{hewitt:2016,
  title={Flow-induced compaction of a deformable porous medium},
  author={Hewitt, Duncan R and Nijjer, Japinder S and Worster, M Grae and Neufeld, Jerome A},
  journal={Physical Review E},
  volume={93},
  number={2},
  pages={023116},
  year={2016},
  publisher={APS}
}

@article{macminn:2015,
  title={Fluid-driven deformation of a soft granular material},
  author={MacMinn, Christopher W and Dufresne, Eric R and Wettlaufer, John S},
  journal={Physical Review X},
  volume={5},
  number={1},
  pages={011020},
  year={2015},
  publisher={APS}
}

@article{desclaux:2026,
  title={Friction modifies the quasistatic mechanical response of a confined, poroelastic medium},
  author={Desclaux, T{\'e}rence and Cuttle, Callum and MacMinn, Christopher W and Liot, Olivier},
  journal={Journal of Fluid Mechanics},
  volume={1035},
  pages={A1},
  year={2026},
  publisher={Cambridge University Press}
}

@article{lutz:2021,
  title={A method for measuring fluid pressure and solid deformation profiles in uniaxial porous media flows},
  author={Lutz, Tyler and Wilen, Larry and Wettlaufer, John},
  journal={Review of Scientific Instruments},
  volume={92},
  number={2},
  year={2021},
  publisher={AIP Publishing}
}

@article{xu:2024,
  title={Hystereses in flow-induced compression of a poroelastic hydrogel},
  author={Xu, Zelai and Yue, Pengtao and Feng, James J},
  journal={Soft Matter},
  volume={20},
  number={35},
  pages={6940--6951},
  year={2024},
  publisher={The Royal Society of Chemistry}
}

@article{thorpe:2015a,
  title={The interfascicular matrix enables fascicle sliding and recovery in tendon, and behaves more elastically in energy storing tendons},
  author={Thorpe, Chavaunne T and Godinho, Marta SC and Riley, Graham P and Birch, Helen L and Clegg, Peter D and Screen, Hazel RC},
  journal={Journal of the mechanical behavior of biomedical materials},
  volume={52},
  pages={85--94},
  year={2015},
  publisher={Elsevier}
}

@article{peltonen:2013,
	author = {Peltonen, Jussi and Cronin, Neil J. and Stenroth, Lauri and Finni, Taija and Avela, Janne},
	date = {2013/05/08},
	doi = {10.1186/2193-1801-2-212},
	id = {Peltonen2013},
	isbn = {2193-1801},
	journal = {SpringerPlus},
	number = {1},
	pages = {212},
	title = {Viscoelastic properties of the Achilles tendon in vivo},
	url = {https://doi.org/10.1186/2193-1801-2-212},
	volume = {2},
	year = {2013}}

@article{schatzmann:1998,
	author = {Schatzmann, L. and Brunner, P. and St{\"a}ubli, H. U.},
	day = {01},
	doi = {10.1007/s001670050224},
	issn = {1433-7347},
	journal = {Knee Surgery, Sports Traumatology, Arthroscopy},
	month = {Apr},
	number = {1},
	pages = {S56--S61},
	title = {Effect of cyclic preconditioning on the tensile properties of human quadriceps tendons and patellar ligaments},
	url = {https://link.springer.com/content/pdf/10.1007/s001670050224.pdf},
	volume = {6},
	year = {1998}}

@article{fiori:2023,
	author = {Fiori, Matilde and Pramanik, Satyajit and MacMinn, Christopher W},
	journal = {Journal of Fluid Mechanics},
	pages = {A2},
	publisher = {Cambridge University Press},
	title = {Flow and deformation due to periodic loading in a soft porous material},
	volume = {974},
	year = {2023}}

@article{thorpe:2014,
	author = {Thorpe, Chavaunne T and Riley, Graham P and Birch, Helen L and Clegg, Peter D and Screen, Hazel RC},
	journal = {Acta biomaterialia},
	number = {7},
	pages = {3217--3224},
	publisher = {Elsevier},
	title = {Effect of fatigue loading on structure and functional behaviour of fascicles from energy-storing tendons},
	volume = {10},
	year = {2014}}

@article{gatt:2015,
	author = {Gatt, Ruben and Wood, Michelle Vella and Gatt, Alfred and Zarb, Francis and Formosa, Cynthia and Azzopardi, Keith M and Casha, Aaron and Agius, Tonio P and Schembri-Wismayer, Pierre and Attard, Lucienne and others},
	journal = {Acta biomaterialia},
	pages = {201--208},
	publisher = {Elsevier},
	title = {Negative Poisson's ratios in tendons: An unexpected mechanical response},
	volume = {24},
	year = {2015}}

@article{vergari:2011,
	author = {Vergari, Claudio and Pourcelot, Philippe and Holden, Laur{\`e}ne and Ravary-Plumio{\"e}n, B{\'e}rang{\`e}re and Gerard, Guillaume and Laugier, Pascal and Mitton, David and Crevier-Denoix, Nathalie},
	journal = {Journal of biomechanics},
	number = {4},
	pages = {719--724},
	publisher = {Elsevier},
	title = {True stress and Poisson's ratio of tendons during loading},
	volume = {44},
	year = {2011}}

@article{goh:2008,
	author = {Goh, KL and Holmes, DF and Lu, H-Y and Richardson, S and Kadler, KE and Purslow, PP and Wess, Timothy James},
    journal = {Journal of biomechanical engineering},
	title = {Ageing changes in the tensile properties of tendons: influence of collagen fibril volume fraction},
	year = {2008}}

@article{jarvinen:2004,
	author = {J{\"a}rvinen, Tero AH and J{\"a}rvinen, Teppo LN and Kannus, Pekka and J{\'o}zsa, L{\'a}szl{\'o} and J{\"a}rvinen, Markku},
	journal = {Journal of orthopaedic research},
	number = {6},
	pages = {1303--1309},
	publisher = {Wiley Online Library},
	title = {Collagen fibres of the spontaneously ruptured human tendons display decreased thickness and crimp angle},
	volume = {22},
	year = {2004}}

@article{kongsgaard:2010,
	author = {Kongsgaard, Mads and Qvortrup, Klaus and Larsen, Jytte and Aagaard, Per and Doessing, Simon and Hansen, Philip and Kjaer, Michael and Magnusson, S Peter},
	journal = {The American journal of sports medicine},
	number = {4},
	pages = {749--756},
	publisher = {Sage Publications Sage CA: Los Angeles, CA},
	title = {Fibril morphology and tendon mechanical properties in patellar tendinopathy: effects of heavy slow resistance training},
	volume = {38},
	year = {2010}}

@article{helland:2013,
	author = {Helland, Christian and Bojsen-M{\o}ller, Jens and Raastad, Truls and Seynnes, Olivier R and Moltubakk, Marie M and Jakobsen, Vidar and Visnes, H{\aa}vard and Bahr, Roald},
	journal = {British journal of sports medicine},
	number = {13},
	pages = {862--868},
	publisher = {BMJ Publishing Group Ltd and British Association of Sport and Exercise Medicine},
	title = {Mechanical properties of the patellar tendon in elite volleyball players with and without patellar tendinopathy},
	volume = {47},
	year = {2013}}

@article{onambele:2007,
	author = {Onamb{\'e}l{\'e}, Gladys NL and Burgess, Katherine and Pearson, Stephen J},
	journal = {Journal of orthopaedic research},
	number = {12},
	pages = {1635--1642},
	publisher = {Wiley Online Library},
	title = {Gender-specific in vivo measurement of the structural and mechanical properties of the human patellar tendon},
	volume = {25},
	year = {2007}}

@article{couppe:2009,
	author = {Couppe, Christian and Hansen, Philip and Kongsgaard, Mads and Kovanen, Vuokko and Suetta, Charlotte and Aagaard, Per and Kjaer, Michael and Magnusson, S Peter},
	journal = {Journal of applied physiology},
	publisher = {American Physiological Society},
	title = {Mechanical properties and collagen cross-linking of the patellar tendon in old and young men},
	year = {2009}}

@article{bojsen-moller:2007,
	author = {Bojsen-M{\o}ller, Jens and Brogaard, Karsten and Have, MJ and Stryger, HP and Kjaer, M and Aagaard, P and Magnusson, SP},
	journal = {Scandinavian Journal of Medicine \& Science in Sports},
	number = {4},
	pages = {415--421},
	publisher = {Wiley Online Library},
	title = {Passive knee joint range of motion is unrelated to the mechanical properties of the patellar tendon},
	volume = {17},
	year = {2007}}

@article{maganaris:2000,
	author = {Maganaris, Constantinos N and Paul, John P},
	journal = {Journal of biomechanics},
	number = {12},
	pages = {1723--1727},
	publisher = {Elsevier},
	title = {Hysteresis measurements in intact human tendon},
	volume = {33},
	year = {2000}}

@article{schwerdt:1980,
	author = {Schwerdt, H and Constantinesco, A and Chambron, J},
	journal = {Journal of Biomechanics},
	number = {11},
	pages = {913--922},
	publisher = {Elsevier},
	title = {Dynamic viscoelastic behaviour of the human tendon in vitro},
	volume = {13},
	year = {1980}}

@article{hennessy:2025,
	author = {Hennessy, Matthew G and Shearer, Tom and Moore, Axel C},
	journal = {Journal of the Mechanics and Physics of Solids},
	pages = {106350},
	publisher = {Elsevier},
	title = {A microstructure-informed continuum model of transversely isotropic, fibre-reinforced hydrogels},
	year = {2025}}

@article{screen:2004a,
	author = {Screen, HRC and Lee, DA and Bader, DL and Shelton, JC},
	journal = {Proceedings of the Institution of Mechanical Engineers, Part H: Journal of Engineering in Medicine},
	number = {2},
	pages = {109--119},
	publisher = {SAGE Publications Sage UK: London, England},
	title = {An investigation into the effects of the hierarchical structure of tendon fascicles on micromechanical properties},
	volume = {218},
	year = {2004}}

@incollection{thorpe:2015,
	author = {Thorpe, Chavaunne T and Birch, Helen L and Clegg, Peter D and Screen, Hazel RC},
	booktitle = {Tendon regeneration},
	pages = {3--39},
	publisher = {Elsevier},
	title = {Tendon physiology and mechanical behavior: structure--function relationships},
	year = {2015}}

@article{diamant:1972,
	author = {Diamant, J and Keller, Andrew and Baer, ELHM and Litt, M and Arridge, RGC},
	journal = {Proceedings of the Royal Society of London. Series B. Biological Sciences},
	number = {1060},
	pages = {293--315},
	publisher = {The Royal Society London},
	title = {Collagen; ultrastructure and its relation to mechanical properties as a function of ageing},
	volume = {180},
	year = {1972}}

@article{kastelic:1978,
	author = {Kastelic, J. and Galeski, A. and Baer, E.},
	doi = {10.3109/03008207809152283},
	issn = {1607-8438},
	journal = {Connective Tissue Research},
	month = jan,
	number = {1},
	pages = {11--23},
	publisher = {Informa UK Limited},
	title = {The Multicomposite Structure of Tendon},
	url = {http://dx.doi.org/10.3109/03008207809152283},
	volume = {6},
	year = {1978}}

@article{kastelic:1980,
	author = {Kastelic, J and Palley, I and Baer, E},
	journal = {Journal of biomechanics},
	number = {10},
	pages = {887--893},
	publisher = {Elsevier},
	title = {A structural mechanical model for tendon crimping},
	volume = {13},
	year = {1980}}

@article{atkinson:1997,
	author = {Atkinson, TS and Haut, RC and Altiero, NJ},
    journal = {Journal of biomechanical engineering},
	title = {A poroelastic model that predicts some phenomenological responses of ligaments and tendons},
	year = {1997}}

@article{butler:1997,
	author = {Butler, SL and Kohles, SS and Thielke, RJ and Chen, C and Vanderby Jr, R},
	journal = {Medical and Biological Engineering and Computing},
	number = {6},
	pages = {742--746},
	publisher = {Springer},
	title = {Interstitial fluid flow in tendons or ligaments: a porous medium finite element simulation},
	volume = {35},
	year = {1997}}

@article{chen:1998,
	author = {Chen, Chih-Tung and Malkus, David S and Vanderby Jr, Ray},
	journal = {Biorheology},
	number = {2},
	pages = {103--118},
	publisher = {Elsevier},
	title = {A fiber matrix model for interstitial fluid flow and permeability in ligaments and tendons},
	volume = {35},
	year = {1998}}

@article{khayyeri:2015,
	author = {Khayyeri, Hanifeh and Gustafsson, Anna and Heuijerjans, Ashley and Matikainen, Marko K and Julkunen, Petro and Eliasson, Pernilla and Aspenberg, Per and Isaksson, Hanna},
	journal = {PloS one},
	number = {6},
	pages = {e0126869},
	publisher = {Public Library of Science San Francisco, CA USA},
	title = {A fibre-reinforced poroviscoelastic model accurately describes the biomechanical behaviour of the rat Achilles tendon},
	volume = {10},
	year = {2015}}

@article{godard:2025,
	author = {Godard, Zoe Charlotte and Moulton, Derek E and Waters, Sarah L},
	journal = {Soft Matter},
	publisher = {Royal Society of Chemistry},
	title = {Cyclic loading of a heterogeneous non-linear poroelastic material},
	year = {2025}}

@article{shearer:2015a,
	author = {Tom Shearer},
	doi = {https://doi.org/10.1016/j.jbiomech.2014.11.031},
	issn = {0021-9290},
	journal = {Journal of Biomechanics},
	number = {2},
	pages = {290-297},
	title = {A new strain energy function for the hyperelastic modelling of ligaments and tendons based on fascicle microstructure},
	url = {https://www.sciencedirect.com/science/article/pii/S0021929014006198},
	volume = {48},
	year = {2015}}

@book{cheng:2016,
	author = {Cheng, Alexander H-D},
	publisher = {Springer},
	title = {Poroelasticity},
	volume = {27},
	year = {2016}}

@article{ciarletta:2008,
	author = {Ciarletta, Pasquale and Dario, Paolo and Micera, Silvestro},
	journal = {Biomaterials},
	number = {6},
	pages = {764--770},
	publisher = {Elsevier},
	title = {Pseudo-hyperelastic model of tendon hysteresis from adaptive recruitment of collagen type I fibrils},
	volume = {29},
	year = {2008}}

@book{coussy:2004,
	author = {Coussy, Olivier},
	publisher = {John Wiley \& Sons},
	title = {Poromechanics},
	year = {2004}}

@article{ahmadzadeh:2015,
	author = {Ahmadzadeh, Hossein and Freedman, Benjamin R and Connizzo, Brianne K and Soslowsky, Louis J and Shenoy, Vivek B},
	journal = {Acta Biomaterialia},
	pages = {83--91},
	publisher = {Elsevier},
	title = {Micromechanical poroelastic finite element and shear-lag models of tendon predict large strain dependent Poisson's ratios and fluid expulsion under tensile loading},
	volume = {22},
	year = {2015}}

@article{wren:2000,
	author = {Wren, Tishya AL and Beaupre, Gary S and Carter, Dennis R},
	journal = {Journal of rehabilitation research and development},
	number = {2},
	pages = {135--144},
	publisher = {REHIBILITATION RESEARCH \& DEVELOPMENT SERVICE},
	title = {Mechanobiology of tendon adaptation to compressive loading through fibrocartilaginous metaplasia},
	volume = {37},
	year = {2000}}

@article{screen:2004,
	author = {Screen, HRC and Bader, DL and Lee, DA and Shelton, JC},
	journal = {Strain},
	number = {4},
	pages = {157--163},
	publisher = {Wiley Online Library},
	title = {Local strain measurement within tendon},
	volume = {40},
	year = {2004}}

@article{rees:2002,
	author = {Rees, Sarah G and Davies, Janet R and Tudor, Debbie and Flannery, Carl R and Hughes, Clare E and Dent, Colin M and Caterson, Bruce},
	journal = {Matrix Biology},
	number = {7},
	pages = {593--602},
	publisher = {Elsevier},
	title = {Immunolocalisation and expression of proteoglycan 4 (cartilage superficial zone proteoglycan) in tendon},
	volume = {21},
	year = {2002}}

@article{safa:2020,
	author = {Safa, Babak N and Bloom, Ellen T and Lee, Andrea H and Santare, Michael H and Elliott, Dawn M},
	journal = {Journal of biomechanics},
	pages = {109892},
	publisher = {Elsevier},
	title = {Evaluation of transverse poroelastic mechanics of tendon using osmotic loading and biphasic mixture finite element modeling},
	volume = {109},
	year = {2020}}

@article{lavagnino:2008,
	author = {Lavagnino, Michael and Arnoczky, Steven P. and Kepich, Eugene and Caballero, Oscar and Haut, Roger C.},
	day = {01},
	doi = {10.1007/s10237-007-0104-z},
	issn = {1617-7940},
	journal = {Biomechanics and Modeling in Mechanobiology},
	month = {October},
	number = {5},
	pages = {405--416},
	title = {A finite element model predicts the mechanotransduction response of tendon cells to cyclic tensile loading},
	url = {https://link.springer.com/content/pdf/10.1007/s10237-007-0104-z.pdf},
	volume = {7},
	year = {2008}}

@article{macminn:2016,
	author = {Christopher W. MacMinn},
	doi = {10.1103/PhysRevApplied.5.044020},
	journal = {Physical Review Applied},
	number = {4},
	title = {Large Deformations of a Soft Porous Material},
	url = {https://doi.org/10.1103/PhysRevApplied.5.044020},
	volume = {5},
	year = {2016}}

%USE THE BELOW OPTIONS IN CASE YOU NEED AUTHOR YEAR FORMAT.
%\bibliographystyle{abbrvnat}
%\bibliography{reference}

\end{document}